\documentclass[aps]{revtex4-2}
\usepackage{float}
\usepackage{eurosym}
\usepackage{graphicx}
\usepackage[colorlinks=true, pdfstartview=FitV, linkcolor=blue, citecolor=red, urlcolor=magenta]{hyperref}
\usepackage{graphicx}
\usepackage{latexsym}
\usepackage{amsmath}
\usepackage{amsfonts}
\usepackage{amssymb}
\usepackage{verbatim}
\usepackage{braket}
\usepackage{extarrows}
\usepackage{hyperref}
\usepackage{orcidlink}
\usepackage{geometry}
\usepackage[english]{babel}
\usepackage{tikz}
\usepackage{xcolor}
\usetikzlibrary{arrows.meta, decorations.pathmorphing, 3d,calc}

\begin{document}

\title{Vacuum Energy and Topological Mass in Interacting Elko and Scalar
Field Theories}
\author{$^{1}$A. J. D. Farias Junior~\orcidlink{0000-0001-5480-546X}}
\email{antonio.farias@ifal.edu.br}
\author{$^{2}$A. Smirnov}
\email{smirnov@ufs.br}
\author{$^{3}$Herondy F. Santana Mota~\orcidlink{0000-0002-7470-1550}}
\email{hmota@fisica.ufpb.br}
\author{$^{3}$E. R. Bezerra de Mello~\orcidlink{0000-0002-6115-5052}}
\email{emello@fisica.ufpb.br}
\affiliation{$^{1}$Instituto Federal de Alagoas,\\
CEP 57466-034, Piranhas, Alagoas, Brazil}

\affiliation{$^{2}$Departamento de F\'{\i}sica, Universidade Federal de Sergipe,\\
CEP 49107-230, S\~{a}o Cr\'{\i}stov\~{a}o, Sergipe, Brazil}

\affiliation{$^{3}$Departamento de F\'{\i}sica, Universidade Federal da Para\'{\i}ba,\\
Caixa Postal 5008, Jo\~{a}o Pessoa, Para\'{\i}ba, Brazil}

\begin{abstract}
In this paper, we consider a four-dimensional system composed of a mass-dimension-one fermionic field, also known as Elko, interacting with a real scalar field. Our main objective is to analyze the Casimir effects associated with this system, assuming that both the Elko and scalar fields satisfy Dirichlet boundary conditions on two large parallel plates separated by a distance $L$. In this scenario, we calculate the vacuum energy density and its first-order correction in the coupling constants of the theory. Additionally, we consider the mass correction for each field separately, namely the topological mass that arises from the boundary conditions imposed on the fields and which also depends on the coupling constants. To develop this analysis, we use the mathematical formalism known as the effective potential, expressed as a path integral in quantum field theory.
\end{abstract}

\maketitle


\section{Introduction}

One of the most important results in theoretical physics was obtained by
H.~Casimir in 1948~\cite{casimir1948attraction}. Casimir proposed that two
large, neutral, conducting, parallel plates are subjected to an attractive
force. The theoretical description of this effect lies within the framework
of quantum field theory, specifically the quantized electromagnetic field.
The boundary conditions imposed on the plates modify the vacuum
fluctuations, giving rise to the attractive force between them. The first
experimental attempt to detect the Casimir effect was carried out by
Sparnaay in 1958~\cite{Sparnaay:1958wg}, but it failed to accurately confirm
the phenomenon due to insufficient precision. Several decades later, the
Casimir effect was successfully observed and measured in a series of
high-accuracy experiments, both for parallel plates~\cite%
{bressi2002measurement} and for other geometries~\cite%
{lamoreaux1997demonstration,lamoreaux1998erratum,
mohideen1998precision,mostepanenko2000new,kim2008anomalies,wei2010results}.
From a theoretical perspective, it has been generalized to different quantum
fields, geometries, boundary conditions, and even curved spacetimes,
becoming a valuable tool for probing fundamental aspects of quantum field
theory and for potential applications in nanotechnology, condensed matter
systems, and cosmology~\cite%
{bordag2001new,mostepanenko1997casimir,brevik2002entropy,
zhang2015thermal,henke2018quantum,milton2019casimir}. Comprehensive reviews
of the Casimir effect and its generalizations can be found in Refs.~\cite%
{bordag2009advances,milton2001casimir,mostepanenko1997casimir}.

Beyond the electromagnetic case, it is now well established that scalar and
fermionic fields also develop Casimir-like effects. Scalar fields under
Robin or helix boundary conditions, or in noncommutative backgrounds,
display rich vacuum structures and thermal corrections~\cite%
{romeo2002casimir, aleixo2021thermal,Escobar:2023hzz}. Fermionic fields, on
the other hand, present additional subtleties related to their anticommuting
nature and boundary conditions, as in the MIT bag model or in spacetimes
with nontrivial topology~\cite{Saghian:2012zy, flachi2017sign,
Saharian:2003sp}. Among these, an especially interesting case is the
so-called Elko field, a fermionic quantum field with mass dimension one. In
this scenario, the Casimir effect was considered in Refs.~\cite%
{Pereira:2016muy, maluf2020casimir}.

Elko fields were first introduced by Ahluwalia and Grumiller~\cite%
{Ahluwalia:2004sz} as eigenspinors of the charge conjugation operator.
Unlike Dirac and Majorana spinors, Elko spinors belong to a non-standard
Wigner class of representations, possessing unusual transformation
properties under the Lorentz group. In particular, they transform under a
non-local representation of Lorentz symmetry, which preserves locality only
along a preferred axis. This peculiar structure forces the Elko field to
have canonical mass dimension one, rather than the usual $3/2$ for fermions,
which drastically restricts the types of renormalizable interactions it may
admit. In fact, the Elko field can only couple renormalizably to scalar
degrees of freedom, most notably the Higgs field, and possibly to gravity.
These features render it largely invisible to Standard Model gauge
interactions, making Elko a natural dark matter candidate.

From the perspective of high-energy physics and cosmology, Elko fields have
been extensively discussed as components of dark matter models, due to their
suppressed couplings to ordinary matter and gauge bosons. They have also
been considered in inflationary scenarios, as drivers of accelerated
expansion, and in models of late-time cosmic acceleration where Elko
condensates mimic dark energy. 
These applications highlight their versatility and justify the study of
Elko in diverse quantum and cosmological settings~\cite{Ahluwalia:2004ab,
Ahluwalia:2004sz, Pereira:2016muy,maluf2020casimir}. In particular,
Casimir-like effects associated with Elko fields offer a way to explore how
exotic spinors interact with boundaries and topology \cite{Pereira:2016muy,
maluf2020casimir}, thereby linking fundamental field-theoretical properties
with potentially observable macroscopic phenomena.

A natural step forward is to investigate Elko fields in interaction with
scalar fields. From the viewpoint of renormalizable Quantum Field Theory
(QFT), this is the most viable scenario, since Elko can couple quadratically
to a scalar field through a dimensionless coupling constant. In the context
of particle physics, such an interaction resonates with the Higgs mechanism,
suggesting possible consequences for mass generation and for dark matter
phenomenology. From the perspective of effective field theory, including
both a quadratic Elko-scalar coupling and a quartic scalar self-interaction
is the minimal and consistent extension to study nontrivial quantum
corrections such as loop effects and induced topological masses.

In order to capture boundary effects, we impose Dirichlet boundary
conditions on both the Elko and the real scalar field. Physically, this
corresponds to confining the fields between two parallel, perfectly
reflecting plates separated by a distance $L$. Such conditions fix the
allowed mode spectrum, directly modifying the vacuum fluctuations and
thereby shaping the vacuum energy. Dirichlet boundary conditions are
particularly well-suited for analytical progress, and they provide a natural
comparison with the electromagnetic Casimir effect in idealized geometries.
Moreover, by enforcing the same boundary conditions on both fields, we
ensure a consistent framework to analyze interaction effects in the vacuum
energy and in the generation of topological masses.

In this work, we analyze a four-dimensional system composed of an Elko
fermionic field interacting with a real scalar field through a quadratic
coupling, while also including a quartic self-interaction for the scalar
sector. Both fields are subjected to Dirichlet boundary conditions on two
large, parallel plates separated by a distance $L$. Employing the path
integral formalism and the effective potential method~\cite%
{jackiw1974functional}, we compute the vacuum energy density (Casimir-like
effect), its first-order interaction corrections, and the topological masses
induced in each field by the combined effect of interactions and boundary
conditions. In particular, we show that the Elko contribution to the vacuum
energy is enhanced relative to the scalar case, and that first-order
coupling corrections give rise to nontrivial boundary-induced mass shifts.
These results extend previous analyses of non-interacting fields~\cite%
{Pereira:2016muy, maluf2020casimir} and highlight how exotic fermions with
suppressed Standard Model interactions can nonetheless manifest observable
quantum vacuum effects.

It is worth mentioning that in the scalar sector the topological mass
correction may turn negative, which signals the possibility of vacuum
instability or symmetry breaking induced by the boundaries. In such
situations, a more refined vacuum stability analysis would be required in
order to determine consistent and physically acceptable values for the
effective scalar mass. This type of analysis naturally connects with the
existence of other possible vacua of the theory, and may reveal
boundary-driven phase transitions between distinct vacuum configurations.

Finally, we note that the effective potential in the present model depends
simultaneously on three background variables, associated with the Elko field
bilinear, the scalar field, and their mutual interaction. This richer
structure increases the possibility of multiple competing vacua, which may
correspond to distinct phases of the theory. While a detailed exploration of
such a vacuum landscape is beyond the scope of the present work, its
existence could have important implications: in some regimes the true ground
state may differ from the perturbative one, or metastable vacua may arise,
potentially leading to boundary-induced phase transitions. Understanding
whether these additional minima carry physical consequences, or are merely
artifacts of the approximation scheme, requires a systematic vacuum
stability analysis, which we leave for future investigation.

The structure of the paper is as follows. In Sec.~\ref{sec2} we introduce
the model, define the interacting Lagrangian, and set up the effective
potential. In Sec.~\ref{sec3} we compute the one-loop corrections using the
generalized zeta-function regularization. Sec.~\ref{sec4} is devoted to the
evaluation of the vacuum energy per unit area of the plates, including
first-order coupling-constant corrections and the emergence of topological
masses. Finally, in Sec.~\ref{sec5} we present our concluding remarks and
perspectives for future work.

Through this paper we use natural units in which both the Planck constant
and the speed of light are set equal to one, $\hbar= c = 1$. 

\section{Effective potential of the Elko field interacting with a real
scalar field}

\label{sec2} 
We consider a system composed of a spin-$1/2$ fermionic field with mass
dimension one, known as the Elko field, interacting (for simplicity) with a
real scalar field. The Euclidean action of the Elko field, including also a
Elko-scalar field interaction, is given by \cite{Ahluwalia:2004ab}: 
\begin{equation}
S\left( \bar{\eta},\eta ,\varphi \right) = S_{\mathrm{E}} + S_{\mathrm{R}} +
S_{\mathrm{int}}\,,  \label{ac1}
\end{equation}
where $S_{\mathrm{E}}$ denotes the action of the free Elko field, 
\begin{equation}
S_{\mathrm{E}}\left( \bar{\eta},\eta \right) = -\int d^{4}x \left( - \bar{%
\eta} \, \square \, \eta + m_{\mathrm{E}}^2 \, \bar{\eta}\eta \right)\,,
\qquad \square = \partial_t^2 + \nabla^2\,.
\end{equation}
Here, $\eta$ represents the Elko field, $\bar{\eta}$ its conjugate, and $m_{%
\mathrm{E}}$ is its mass.

The term $S_{\mathrm{R}}$ corresponds to the action of the scalar field,
including the quartic self-interaction: 
\begin{equation}
S_{\mathrm{R}}(\varphi) = -\frac{1}{2} \int d^4x \left( -\varphi \, \square
\, \varphi + m_{\mathrm{R}}^2 \, \varphi^2 \right) - \int d^4x \, \frac{%
\lambda_\varphi}{4!} \, \varphi^4\,,
\end{equation}
where $\varphi$ represents the real scalar field, $m_{\mathrm{R}}$ its mass,
and $\lambda_\varphi$ the quartic self-coupling constant.

Finally, the Elko-scalar field interaction action is defined as \cite%
{Ahluwalia:2004ab}: 
\begin{equation}
S_{\mathrm{int}}\left( \bar{\eta},\eta ,\varphi \right) = - \int d^4x \, g
\, \varphi^2 \, \bar{\eta}\eta \,,
\end{equation}
where $g$ is the coupling constant characterizing the interaction between
the Elko and scalar fields.

For the system described above, the vacuum-to-vacuum transition function is
written as 
\begin{equation}
Z\left( \bar{\eta},\eta ,\varphi \right) = \int \mathcal{D}\bar{\eta} \, 
\mathcal{D}\eta \, \mathcal{D}\varphi \, \exp \left( S_{\mathrm{E}} + S_{%
\mathrm{R}} + S_{\mathrm{int}} \right) \,.
\end{equation}

In order to construct the effective potential, we use the path integral
approach. In this paper, we follow the procedure detailed in Refs.~\cite%
{jackiw1974functional, ryder1996quantum, toms1980symmetry} (see also \cite%
{cruz2020casimir, porfirio2021ground, PhysRevD.107.125019}). For the
reader's convenience, we present only the main steps.

We shift the fields around fixed background fields, i.e., 
\begin{equation}
\eta = \Psi + \chi \,, \qquad \varphi = \Phi + \rho \,,
\end{equation}
where $\chi$ and $\rho$ represent quantum fluctuations, and $\Psi$ and $\Phi$
are fixed classical background fields. Note that $\Phi$ is a scalar field, while $\Psi$ is a single Grassmann variable, reflecting the fermionic nature of the Elko field. The procedure above is analogous to what
was done in Ref.~\cite{PhysRevD.107.125019}. In contrast, in the present
case, due to the fermionic nature of the Elko field, there is no need to set 
$\Phi = 0$ in advance, unlike in \cite{PhysRevD.107.125019}.

The effective potential can then be expanded as 
\begin{equation}
V_{\mathrm{eff}}\left( \bar{\Psi}, \Psi, \Phi \right) = V^{(0)}\left( \bar{%
\Psi}, \Psi, \Phi \right) + V^{(1)}\left( \bar{\Psi}, \Psi, \Phi \right) +
V^{(2)}\left( \bar{\Psi}, \Psi, \Phi \right) \,,  \label{rc2.1}
\end{equation}
where $V^{(0)}$ is the classical (tree-level) contribution: 
\begin{equation}
V^{(0)}\left( \bar{\Psi}, \Psi, \Phi \right) = m_{\mathrm{E}}^2 \, \bar{\Psi}
\Psi + m_{\mathrm{R}}^2 \, \Phi^2 + \frac{\lambda_\varphi}{4!} \Phi^4 + g \,
\Phi^2 \bar{\Psi} \Psi + \xi \left( \bar{\Psi}, \Psi, \Phi, C_i \right) \,,
\label{v0}
\end{equation}
with 
\begin{equation}
\xi \left( \bar{\Psi}, \Psi, \Phi, C_i \right) = \frac{C_1}{4!} \Phi^4 + 
\frac{C_2}{2} \Phi^2 + C_3 + C_4 \bar{\Psi} \Psi + C_5 \Phi^2 \bar{\Psi}
\Psi \,,  \label{fr01}
\end{equation}
representing the renormalization contributions with constants $C_i$ $%
(i=1,\dots,5)$.

The one-loop contribution $V^{(1)}$ can be expressed as a path integral \cite%
{toms1980interacting, PhysRevD.107.125019}, that is, 
\begin{equation}
V^{(1)}\left( \bar{\Psi}, \Psi, \Phi \right) = - \frac{1}{\Omega_4} \ln \int 
\mathcal{D}\bar{\eta} \, \mathcal{D}\eta \, \mathcal{D}\varphi \, \exp %
\Bigg[ - \int d^4x \, \bar{\eta} \hat{A} \eta - \frac{1}{2} \int d^4x \,
\varphi \hat{B} \varphi \Bigg] \,,  \label{v1}
\end{equation}
where $\Omega_4$ is the four-dimensional spacetime volume and the operators $%
\hat{A}$ and $\hat{B}$ are defined as 
\begin{align}
\hat{A} &= \left( -\square + m_{\mathrm{E}}^2 + g \Phi^2 \right) \mathbf{1}%
_{4\times4} \,,  \notag \\
\hat{B} &= -\square + m_{\mathrm{R}}^2 + \frac{\lambda_\varphi}{2} \Phi^2 +
2 g \bar{\Psi} \Psi \,,  \label{d22}
\end{align}
with $\mathbf{1}_{4\times4}$ denoting the $4\times 4$ identity matrix.

The two-loop contribution $V^{(2)}$ is most conveniently computed using the
Feynman diagrams of the theory under consideration, which will be done in
Sec.~\ref{sec4}.

Given the form of the one-loop correction in Eq.~(\ref{v1}), we can separate
the contributions from each field in the path integral. The path integral
associated with the real scalar field has been written in a convenient form
in several references \cite{PhysRevD.107.125019, toms1980interacting,
porfirio2021ground}. Now, for the Elko field we write 
\begin{equation}
V_{\mathrm{E}}^{(1)} = - \frac{1}{\Omega_4} \ln \int \mathcal{D}\bar{\eta} 
\mathcal{D}\eta \, \exp \Big( - \int d^4x \, \bar{\eta} \hat{A} \eta \Big) %
\,.
\end{equation}

To write $V_{\mathrm{E}}^{(1)}$ we follow the steps described in \cite%
{PhysRevD.110.045006}, although some of these steps are indicated here.
First, we assume that the operator $\hat{A}$ has a complete set of
eigenfunctions and eigenvalues, i.e., 
\begin{equation}
\hat{A} \psi_j = \Lambda_j \psi_j \,, \qquad \int d^4x \, \psi_i^\dagger(x)
\psi_j(x) = \delta_{ij} \,.
\end{equation}

Since $\{\psi_i\}$ forms a complete basis of spinors, we can expand the Elko
field as 
\begin{equation}
\eta(x) = \sum_i \chi_i \psi_i(x) \,, \qquad \bar{\eta}(x) = \sum_i \bar{\chi%
}_i \psi_i^\dagger(x) \,,
\end{equation}
where $\chi_i$ and $\bar{\chi}_i$ are independent Grassmann variables.
Therefore, $V_{\mathrm{E}}^{(1)}$ becomes 
\begin{equation}
V_{\mathrm{E}}^{(1)} = - \frac{1}{\Omega_4} \ln \int \prod_j \frac{d\bar{\chi%
}_j d\chi_j}{\alpha^2} \, \exp \Big( - \sum_j \bar{\chi}_j \chi_j \Lambda_j %
\Big) \,,  \label{v1.01}
\end{equation}
where the mass-dimensional constant $\alpha$ is an integration measure in
the functional space of the Elko field. Expanding the exponential and using
the properties of Grassmann integrals, one finds 
\begin{equation}
V_{\mathrm{E}}^{(1)} = - \frac{1}{\Omega_4} \ln \prod_j \frac{\Lambda_j}{%
\alpha^2} = - \frac{1}{\Omega_4} \ln \det \frac{\hat{A}}{\alpha^2} \,.
\end{equation}

Using the well-known relation between determinants and traces, $\det M =
\exp(\mathrm{tr} \ln M)$, we can rewrite this as 
\begin{equation}
V_{\mathrm{E}}^{(1)} = - \frac{1}{\Omega_4} \mathrm{tr} \ln \frac{\hat{A}}{%
\alpha^2} = - \frac{4}{\Omega_4} \sum_i \ln \frac{\Lambda_i}{\alpha^2} \,,
\end{equation}
where the factor of $4$ comes from the trace of the $4 \times 4$ identity
matrix.

In Ref.~\cite{Ahluwalia:2004ab}, a $\phi$-dependent matrix, $\mathcal{G}%
(\phi)$, is introduced and defined in terms of spin-sum results. It
satisfies the property $\mathcal{G}(\phi) = -\mathcal{G}(\phi + \pi)$ and
contributes to the system only if a preferred direction exists. In the
present setup, however, the contribution of $\mathcal{G}(\phi)$ vanishes due
to the axial symmetry around the $z$-axis and the fact that $\int_0^{2\pi}
d\phi\, \mathrm{tr}\,\mathcal{G}(\phi) = 0$. Nevertheless, nontrivial
effects of $\mathcal{G}(\phi)$ could, in principle, arise in situations
where the azimuthal symmetry is explicitly broken. Examples include (i)
anisotropic or structured plates that single out a preferred direction in
the $xy$-plane, (ii) the presence of external vector backgrounds that couple
differently to distinct azimuthal angles, or (iii) the analysis of
directional correlation functions without performing the full angular
integration. In such cases, the Fourier zero mode of $\mathcal{G}(\phi)$
would survive and could potentially modify the spectral sums and loop
corrections.

Introducing the generalized zeta function \cite{toms1980symmetry,
hawking1977zeta, PhysRevD.107.125019} constructed from the eigenvalues $%
\Lambda_j$, we can write 
\begin{equation}
V_{\mathrm{E}}^{(1)} = \frac{4}{\Omega_4} \Big[ \zeta_{\eta}^{\prime }(0) +
\zeta_{\eta}(0) \ln \alpha^2 \Big] \,,  \label{ve1}
\end{equation}
where 
\begin{equation}
\zeta(s) = \sum_{\sigma} \Lambda_{\sigma}^{-s},  \label{z1}
\end{equation}
is the generalized zeta function, which converges for Re$(s)>2$ and is
regular at $s= 0$, admitting an analytic continuation to other values of $s$%
. Note that $\sigma$ represents the set of quantum numbers of the system.

Similarly, for the scalar field we obtain \cite{PhysRevD.107.125019,
toms1980interacting, porfirio2021ground}: 
\begin{align}
V_{\mathrm{R}}^{(1)} &= - \frac{1}{\Omega_4} \ln \int \mathcal{D}\varphi
\exp \Big( - \frac{1}{2} \int d^4x \, \varphi \hat{B} \varphi \Big)  \notag
\\
&= - \frac{1}{2\Omega_4} \Big[ \zeta_{\varphi}^{\prime }(0) +
\zeta_{\varphi}(0) \ln \beta^2 \Big] \,.  \label{vh1}
\end{align}
Here, $\beta$ is also a mass-dimensional constant that serves as an
integration measure in the functional space of the scalar field. Both $%
\alpha $ and $\beta$ are auxiliary parameters that will be removed by
imposing appropriate renormalization conditions on the fields.

Collecting the results in Eqs.~\eqref{ve1} and \eqref{vh1}, the one-loop
correction to the effective potential is 
\begin{align}
V^{(1)}(\bar{\Psi}, \Psi, \Phi) &= V_{\mathrm{E}}^{(1)}(\bar{\Psi}, \Psi,
\Phi) + V_{\mathrm{R}}^{(1)}(\bar{\Psi}, \Psi, \Phi).  \label{lc1}
\end{align}
Comparing Eqs.~(\ref{ve1}) and (\ref{vh1}), one observes that the
contribution from the Elko field is eight times larger than that of the real
scalar field. This result contrasts with the standard Dirac fermion, whose
contribution is four times larger~\cite{venancio2024thermal}. Furthermore,
our finding also differs from that of Ref.~\cite{Pereira:2018xyl}, where the
authors analyzed the partition function associated with the Elko field.

In analogy with the analysis presented in Ref.~\cite{Ahluwalia:2004ab},
where the interaction between the Elko field and the Higgs field plays a
central phenomenological role, we focus here on the coupling between the
Elko field and a real scalar field. This choice is made for the sake of
simplicity, allowing a clearer examination of the vacuum energy corrections
induced by the scalar-fermionic interaction, while avoiding the additional
complications arising from gauge structure or spontaneous symmetry breaking.
Nevertheless, the formulation and results presented in this work can be
straightforwardly generalized to the case of a complex scalar field-such as
the Higgs doublet-since the essential features of the interaction and its
impact on the effective potential remain formally analogous.

The two-loop contribution $V^{(2)}(\bar{\Psi}, \Psi, \Phi)$, with the help
of Feynman diagrams, can also be expressed in terms of generalized zeta
functions if one is only interested in vacuum contributions \cite%
{cruz2020casimir, porfirio2021ground}. Its explicit form will be presented
in Sec.~\ref{sec4}.


\section{Generalized zeta function and one-loop correction}

\label{sec3} 
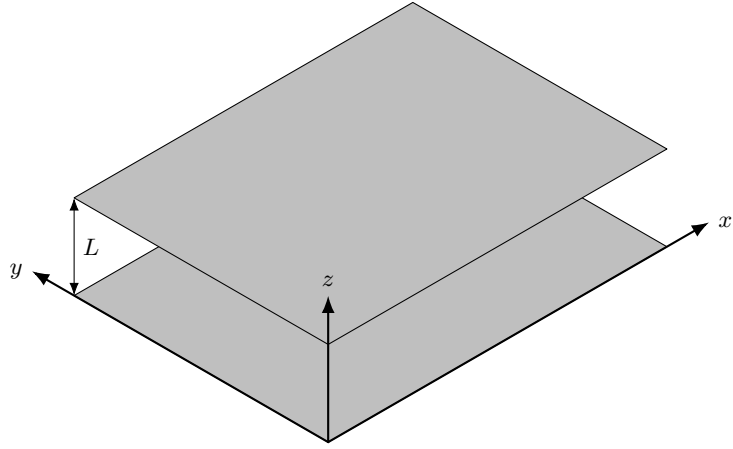
\begin{figure}[h]
\centering
\begin{tikzpicture}[scale=1.3, line join=round, line cap=round, >=Latex,
                    x={(0.866cm,0.5cm)}, y={( -0.866cm,0.5cm)}, z={(0cm,1cm)}]
  \def\Lx{4}   
  \def\Ly{3}   
  \def\zsep{1} 

  \fill[gray!50] (0,0,0) -- (\Lx,0,0) -- (\Lx,\Ly,0) -- (0,\Ly,0) -- cycle;
  \draw[black] (0,0,0) -- (\Lx,0,0) -- (\Lx,\Ly,0) -- (0,\Ly,0) -- cycle;

  \fill[gray!50] (0,0,\zsep) -- (\Lx,0,\zsep) -- (\Lx,\Ly,\zsep) -- (0,\Ly,\zsep) -- cycle;
  \draw[black] (0,0,\zsep) -- (\Lx,0,\zsep) -- (\Lx,\Ly,\zsep) -- (0,\Ly,\zsep) -- cycle;

  \draw[->,thick] (0,0,0) -- (4.5,0,0) node[right] {$x$};
  \draw[->,thick] (0,0,0) -- (0,3.5,0) node[left] {$y$};
  \draw[->,thick] (0,0,0) -- (0,0,1.5) node[above] {$z$};

\draw[<->, thin, black] 
    (0, \Ly, 0) -- node[right] {$L$} (0, \Ly, \zsep);

\end{tikzpicture}
\caption{A schematic representation of two perfectly reflecting parallel
plates lying in the $x$-$y$ plane and separated by a distance $L$ along the $%
z$-axis is shown. Both the Elko and the real scalar fields satisfy Dirichlet
boundary conditions at the plates, located at $z=0$ and $z=L$. }
\label{fig1}
\end{figure}

In this section, we employ the generalized zeta-function technique to obtain
the one-loop correction to the effective potential of the theory under
consideration. The fields satisfy Dirichlet boundary conditions imposed on
two large, perfectly reflecting parallel plates separated by a distance $L$
along the $z$-axis (see Fig.~\ref{fig1}). 


\subsection{Dirichlet boundary conditions for the Elko and scalar fields}

In Ref.~\cite{maluf2020casimir}, the authors analyzed the Casimir energy
associated with an Elko field, assuming that the field satisfies Dirichlet
boundary conditions on two large plates separated by a distance $L$. In this
section, we revisit this problem by considering a more general situation, in
which the Elko field is coupled to a real scalar field. The latter, in turn,
is assumed to satisfy the same boundary conditions as the Elko field.


We now want to calculate the one-loop effective potential associated with
the Elko field. In the case of Dirichlet boundary conditions, the Elko field 
$\eta$ and its conjugate $\bar{\eta}$ satisfy the following relation at the
plates \cite{Pereira:2016muy, maluf2020casimir}: 
\begin{equation}
\eta(\tau,x,y,0) = \eta(\tau,x,y,L) = 0 \,,  \label{l1}
\end{equation}
This plays a crucial role in the evaluation of the effective potential, as
it determines the allowed mode spectrum of the field between the boundaries,
thereby affecting the summation over quantum fluctuations that contributes
to the one-loop corrections.

Therefore, under the above condition, the eigenvalues of the operator $\hat{A%
}$, presented in Eq.~(\ref{d22}), take the form given by 
\begin{equation}
\Lambda_{\sigma} = k_{\tau}^2 + k_x^2 + k_y^2 + \left( \frac{n \pi}{L}
\right)^2 + M_{\mathrm{E}}^2\,, \quad n \in \mathbb{Z}_{+}^{\ast}\,,
\label{eigenf}
\end{equation}
where the index $\sigma$ denotes the set of quantum numbers $\left(
k_{\tau}, k_{x}, k_{y}, n \right)$ and we have defined the quantity $M_{%
\mathrm{E}}^{2}$ as 
\begin{equation}
M_{\mathrm{E}}^{2} = m_{\mathrm{E}}^{2} + g \Phi^2 \,.  \label{me}
\end{equation}

In addition, in the case under consideration, the background field $\Phi$
can be taken as non-zero. This will not be the case when dealing with two
scalar fields, as in Refs.~\cite{PhysRevD.107.125019, toms1980interacting,
doi:10.1142/S021827182450069X}.

The scalar field also obeys Dirichlet boundary conditions, and therefore
satisfies a relation analogous to Eq.~(\ref{l1}). Hence, the eigenvalues of
the operator $\hat{B}$, Eq.~(\ref{d22}), can be written as 
\begin{equation}
\Lambda_{\sigma} = k_{\tau}^{2} + k_{x}^{2} + k_{y}^{2} + \left( \frac{j \pi%
}{L} \right)^2 + M_{g}^{2} \,, \quad j \in \mathbb{Z}_{+}^{\ast} \,,
\label{rc16.1}
\end{equation}
where the quantity $M_{g}^{2}$ is defined as 
\begin{equation}
M_{g}^{2} = m_{\mathrm{R}}^{2} + \frac{\lambda_{\varphi}}{2} \Phi^2 + 2 g 
\bar{\Psi} \Psi \,.  \label{mg}
\end{equation}
Here, $\sigma$ denotes the set of quantum numbers $\left( k_{\tau}, k_{x},
k_{y}, j \right)$. Note that in both cases, namely Eqs.~\eqref{eigenf} and %
\eqref{rc16.1}, the momenta $\left( k_{\tau}, k_{x}, k_{y} \right)$ are
continuous, whereas $j$ and $n$ are discrete.



\subsection{Loop Correction and Generalized Zeta Functions}

Knowing the explicit form of the eigenvalues presented in Eqs.~(\ref{eigenf}%
) and (\ref{rc16.1}), we can construct the generalized zeta functions
associated with each field. We begin by considering the generalized zeta
function corresponding to the Elko field. Using the eigenvalues in Eq.~(\ref%
{eigenf}), the generalized zeta function, Eq.~(\ref{z1}), is given by 
\begin{equation}
\zeta _{\mathrm{E}}\left( s\right) =\frac{\Omega _{3}}{\left( 2\pi
\right)^{3}}\int dk_{\tau}dk_{x}dk_{y}\sum_{n=1}^{\infty }\left[
k_{\tau}^{2}+k_{y}^{2}+k_{x}^{2}+\left( \frac{n\pi }{L}\right) ^{2}+M_{%
\mathrm{E}}^{2}\right] ^{-s}\,,
\end{equation}
where $\Omega_3$ is a volume-like parameter associated with the continuous
momenta $(k_{\tau}, k_x, k_y)$.

To rewrite this zeta function in a more convenient form, we follow the
procedure outlined, for instance, in \cite{cruz2020casimir}, highlighting
some steps for clarity. First, we use the identity 
\begin{equation}
w^{-s}=\frac{2}{\Gamma \left( s\right) }\int_{0}^{\infty } d\rho \
\rho^{2s-1} e^{-w\rho ^{2}}\,,
\end{equation}
and perform the resulting Gaussian integrals. This yields an expression
suitable for identifying the integral representation of the gamma function $%
\Gamma \left( z\right)$ \cite{abramowitz1965handbook}, namely 
\begin{equation}
\frac{\Gamma \left( s\right) }{2}=\int_{0}^{\infty } d\mu \ \mu^{2s -1}
e^{-\mu ^{2}}\,.
\end{equation}

Following these steps, the generalized zeta function associated with the
Elko field can be expressed as 
\begin{equation}
\zeta _{\mathrm{E}}\left( s\right) =\frac{\Omega _{4}\pi ^{\frac{3}{2}-2s}}{%
8L^{4-2s}}\frac{\Gamma \left( s-\frac{3}{2}\right) }{\Gamma \left(s\right) }%
\sum_{n=1}^{\infty }\left( n^{2}+\frac{M_{\mathrm{E}}^{2}L^{2}}{\pi ^{2}}%
\right) ^{\frac{3}{2}-s}\,,  \label{zf1}
\end{equation}
where $\Omega _{4}=\Omega _{3}L$. Using the Epstein-Hurwitz zeta function 
\cite{Elizalde:1995hck, Elizalde_hiroshima}, i.e., 
\begin{align}
\sum_{l=1}^\infty \left[l^2 + \kappa^2 \right]^{\frac{3}{2} - s} = & -\frac{%
\kappa^{3 - 2s}}{2} + \frac{\sqrt{\pi}}{2} \frac{\Gamma(s - 2)}{%
\Gamma\left(s - \frac{3}{2}\right)} \kappa^{4 - 2s}  \notag \\
& + \frac{2 \kappa^{2 - s} \pi^{s - \frac{3}{2}}}{\Gamma\left(s - \frac{3}{2}%
\right)} \sum_{j=1}^\infty j^{s-2} K_{s-2}\left(2 \pi j \kappa \right),
\end{align}
where $K_{\gamma }(x)$ denotes the Macdonald function \cite%
{abramowitz1965handbook}, we can perform the sum over $n$ in Eq.~(\ref{zf1})
and rewrite the generalized zeta function as a sum of three terms: 
\begin{eqnarray}
\zeta _{\mathrm{\eta}}\left( s\right) &=&-\frac{\Omega _{4}M_{\mathrm{E}%
}^{3-2s}}{16\pi ^{\frac{3}{2}}L}\frac{\Gamma \left( s-\frac{3}{2}\right) }{%
\Gamma\left( s\right) }+\frac{\Omega _{4}M_{\mathrm{E}}^{4-2s}}{16\pi ^{2}}%
\frac{\Gamma \left( s-2\right) }{\Gamma \left( s\right) }  \notag \\
&&+\frac{\Omega _{4}M_{\mathrm{E}}^{2-s}}{4\pi ^{2}L^{2-s}}\frac{1}{\Gamma
\left( s\right) }\sum_{n=1}^{\infty }n^{s-2}K_{s-2}\left( 2nM_{\mathrm{E}%
}L\right) \,.  \label{ze01}
\end{eqnarray}

Evaluating the zeta function and its derivative in the limit $s\rightarrow 0$%
, we obtain from Eq.~(\ref{lc1}) the one-loop correction to the effective
potential due to the Elko field, i.e., 
\begin{eqnarray}
V_{\mathrm{E}}^{\left( 1\right) }\left( \bar{\Psi},\Psi ,\Phi \right) &=&%
\frac{M_{\mathrm{E}}^{4}}{8\pi ^{2}}\left[ \frac{3}{2}-\ln\left(\frac{M_{%
\mathrm{E}}^{2}}{\alpha ^{2}}\right)\right] -\frac{M_{\mathrm{E}}^{3}}{3\pi L%
}  \notag \\
&&+\frac{M_{\mathrm{E}}^{2}}{\pi ^{2}L^{2}}\sum_{n=1}^{\infty
}n^{-2}K_{2}\left( 2nM_{\mathrm{E}}L\right) \,.  \label{vd}
\end{eqnarray}

Here, the first term on the right-hand side of Eq.~(\ref{vd}) is independent
of the parameter $L$ and therefore does not contribute to the vacuum energy
density. The second term grows with positive powers of the mass and should therefore be discarded. 
This is consistent with a normalization condition requiring that the renormalized vacuum energy vanish in the large-mass limit~\cite{bordag2009advances}. Hence, only the
third term contributes to the renormalized vacuum energy density, as we will
see later.

By a similar calculation, we construct the generalized zeta function
associated with the scalar field. Using the eigenvalues presented in Eq.~(%
\ref{rc16.1}) and following steps analogous to those described above, we
obtain 
\begin{eqnarray}
\zeta _{\mathrm{\varphi}}\left( s\right) &=&\frac{\Omega _{4}M_{g}^{4-2s}}{%
16\pi ^{2}}\frac{\Gamma \left( s-2\right) }{\Gamma \left( s\right) }-\frac{%
\Omega _{4}M_{g}^{3-2s}}{16\pi ^{\frac{3}{2}}L}\frac{\Gamma \left( s-\frac{3%
}{2}\right) }{\Gamma \left( s\right) }  \notag \\
&&+\frac{\Omega _{4}M_{g}^{2-s}}{4\pi ^{2}L^{2-s}}\frac{1}{\Gamma \left(
s\right) }\sum_{j=1}^{\infty }j^{s-2}K_{s-2}\left( 2jM_{g}L\right) \,.
\label{zh01}
\end{eqnarray}

Evaluating this zeta function and its derivative at $s=0$, we find the
one-loop correction to the effective potential due to the scalar field as 
\begin{eqnarray}
V_{\mathrm{R}}^{\left( 1\right) }\left( \bar{\Psi},\Psi ,\Phi \right) &=&%
\frac{M_{g}^{4}}{64\pi ^{2}}\left[ \ln\left( \frac{M_{g}^{2}}{\beta ^{2}}%
\right)-\frac{3}{2}\right] +\frac{M_{g}^{3}}{24\pi L}  \notag \\
&&-\frac{M_{g}^{2}}{8\pi ^{2}L^{2}}\sum_{j=1}^{\infty }j^{-2}K_{2}\left(
2jM_{g}L\right) \,.  \label{vh01}
\end{eqnarray}

Combining the results from Eqs.~(\ref{vd}) and (\ref{vh01}), together with
Eqs.~(\ref{rc2.1}) and (\ref{v0}), we write the non-renormalized effective
potential up to one-loop order as 
\begin{eqnarray}
V_{\text{eff}}\left( \bar{\Psi},\Psi ,\Phi \right) &=& m_{\mathrm{E}}^{2}%
\bar{\Psi}\Psi +\frac{1}{2}m_{\mathrm{R}}^{2}\Phi ^{2}+\frac{%
\lambda_{\varphi }}{4!}\Phi ^{4} + g\Phi ^{2}\bar{\Psi}\Psi  \notag \\
&& +\xi \left( \bar{\Psi},\Psi ,\Phi ,C_{i}\right) +\frac{M_{\mathrm{E}}^{4}%
}{8\pi ^{2}}\left[ \frac{3}{2}-\ln\left( \frac{M_{\mathrm{E}}^{2}}{\alpha
^{2}}\right)\right]  \notag \\
&&+\frac{M_{g}^{4}}{64\pi ^{2}}\left[\ln\left( \frac{M_{g}^{2}}{\beta ^{2}}%
\right)-\frac{3}{2}\right] + \frac{M_{g}^{3}}{24\pi L}-\frac{M_{\mathrm{E}%
}^{3}}{3\pi L}  \notag \\
&&+\frac{M_{\mathrm{E}}^{2}}{\pi ^{2}L^{2}}\sum_{n=1}^{\infty
}n^{-2}K_{2}\left( 2nM_{\mathrm{E}}L\right)  \notag \\
&&-\frac{M_{g}^{2}}{8\pi ^{2}L^{2}}\sum_{j=1}^{\infty }j^{-2}K_{2}\left(
2jM_{g}L\right) \,.  \label{ve01}
\end{eqnarray}
Note that although the effective potential above may in principle exhibit
several minima, we restrict ourselves to the vacuum defined at 
\begin{equation}
v=({\bar{\Psi},\Psi ,\Phi})=(0,0,0).  \label{vacuum}
\end{equation}
This choice is motivated by the fact that it preserves the symmetries of the
theory, corresponds to the trivial background around which the perturbative
expansion is consistently performed, and provides a natural renormalization
point where the physical parameters (masses and couplings) are defined.

Before proceeding to the calculation of the vacuum energy, it is necessary
to renormalize the effective potential in Eq.~(\ref{ve01}). The
renormalization conditions, taken at the vacuum state $v$, are given by 
\begin{equation}  \label{rc14.5}
\begin{aligned} \left.\frac{d^{4}V_{\mathrm{eff}}}{d\Phi ^{4}}\right|_{v} &=
\lambda _{\varphi}, &\qquad \left.\frac{d^{4}V_{\mathrm{eff}}}{d\Phi ^{2}\,
d\Psi \, d\bar{\Psi}}\right|_{v} &= 2g, \end{aligned}
\end{equation}
for the self- and cross-couplings, respectively. These conditions determine
the renormalization constants $C_1$ and $C_5$, according to Eqs.~\eqref{v0}
and \eqref{fr01}.

In addition, to determine the coefficients $C_2$ and $C_4$ one must impose
renormalization conditions for the masses of the real scalar and Elko
fields, respectively, as follows: 
\begin{equation}  \label{rc14.5.2}
\begin{aligned} \left.\frac{d^{2}V_{\mathrm{eff}}}{d\Phi^{2}}\right|_{v} &=
m_{\mathrm{R}}^{2}, &\qquad \left.\frac{d^{2}V_{\mathrm{eff}}}{d\Psi\,
d\bar{\Psi}}\right|_{v} &= m_{\mathrm{E}}^{2}. \end{aligned}
\end{equation}

A final condition is still required to determine the coefficient $C_3$. This
is done by imposing that the effective potential vanishes in the vacuum
state $v$, i.e., 
\begin{equation}  \label{C3}
V_{\mathrm{eff}}\big|_{v}=0.
\end{equation}

Note that the conditions above follow in accordance with Refs.~\cite%
{PhysRevD.107.125019, toms1980interacting, porfirio2021ground}. It is worth
noticing that the second condition in Eq.~\eqref{rc14.5} for $g$ arises as a
consequence of the background field $\Phi$ being nonzero, in contrast to
Refs.~\cite{PhysRevD.107.125019, toms1980interacting, porfirio2021ground}.
Moreover, these conditions are to be applied in the Minkowski spacetime
limit, $L\rightarrow \infty$. Enforcing these conditions determines the
explicit form of the renormalization constants $C_{i}$ in the function $\xi
\!\left( \bar{\Psi}, \Psi , \Phi , C_{i}\right)$ of Eq.~(\ref{fr01}).

The renormalization procedure is detailed in the Appendix~\ref{ren}. Here,
we present the renormalization coefficients and the resulting renormalized
effective potential. The coefficients are written as 
\begin{align}
C_{1} &= \frac{3g^{2}}{\pi^{2}} \ln\left( \frac{m_{\mathrm{E}}^{2}}{%
\alpha^{2}}\right) - \frac{3 \lambda_{\varphi}^{2}}{32 \pi^{2}} \ln\left( 
\frac{m_{\mathrm{R}}^{2}}{\beta^{2}}\right),  \notag \\
C_{2} &= \frac{\lambda_{\varphi} m_{\mathrm{R}}^{2}}{32 \pi^{2}} \left[1 -
\ln\left( \frac{m_{\mathrm{R}}^{2}}{\beta^{2}}\right)\right] + \frac{g m_{%
\mathrm{E}}^{2}}{2 \pi^{2}} \left[ \ln\left( \frac{m_{\mathrm{E}}^{2}}{%
\alpha^{2}}\right) - 1 \right],  \notag \\
C_{3} &= -\frac{m_{\mathrm{E}}^{4}}{8 \pi^{2}} \left[ \frac{3}{2} -
\ln\left( \frac{m_{\mathrm{E}}^{2}}{\alpha^{2}}\right) \right] - \frac{m_{%
\mathrm{R}}^{4}}{64 \pi^{2}} \left[ \ln\left( \frac{m_{\mathrm{R}}^{2}}{%
\beta^{2}}\right) - \frac{3}{2} \right],  \notag \\
C_{4} &= - \frac{g m_{\mathrm{R}}^{2}}{32 \pi^{2}} - \frac{4 m_{\mathrm{R}%
}^{2} g}{54 \pi^{2}} \left[ \ln\left( \frac{m_{\mathrm{R}}^{2}}{\beta^{2}}%
\right) - \frac{3}{2} \right],  \notag \\
C_{5} &= - \frac{g \lambda_{\varphi}}{32 \pi^{2}} \left[ \ln\left( \frac{m_{%
\mathrm{R}}^{2}}{\beta^{2}}\right) - \frac{3}{2} \right].  \label{rcon}
\end{align}

These coefficients yield the renormalized effective potential as follows: 
\begin{equation}
\begin{aligned} V_{\mathrm{eff}}^{\mathrm{ren}}(\bar{\Psi}, \Psi, \Phi) &=
m_{\mathrm{E}}^{2} \bar{\Psi} \Psi + \frac{1}{2} m_{\mathrm{R}}^{2} \Phi^{2}
+ \frac{\lambda_{\varphi}}{4!} \Phi^{4} + g \Phi^{2} \bar{\Psi} \Psi \\
&\quad + \frac{g^{2} \Phi^{4}}{8 \pi^{2}} \left[ \frac{3}{2} - \ln \left(
\frac{M_{\mathrm{E}}^{2}}{m_{\mathrm{E}}^{2}} \right) \right] +
\frac{\lambda_{\varphi}^{2} \Phi^{4}}{256 \pi^{2}} \left[ \ln \left(
\frac{M_{g}^{2}}{m_{\mathrm{R}}^{2}} \right) - \frac{3}{2} \right] \\ &\quad
+ \frac{g m_{\mathrm{E}}^{2} \Phi^{2}}{4 \pi^{2}} \left[ \frac{1}{2} - \ln
\left( \frac{M_{\mathrm{E}}^{2}}{m_{\mathrm{E}}^{2}} \right) \right] +
\frac{\lambda_{\varphi} m_{\mathrm{R}}^{2} \Phi^{2}}{64 \pi^{2}} \left[ \ln
\left( \frac{M_{g}^{2}}{m_{\mathrm{R}}^{2}} \right) - \frac{1}{2} \right] \\
&\quad - \frac{m_{\mathrm{E}}^{4}}{8 \pi^{2}} \ln \left(
\frac{M_{\mathrm{E}}^{2}}{m_{\mathrm{E}}^{2}} \right) +
\frac{m_{\mathrm{R}}^{4}}{64 \pi^{2}} \ln \left( \frac{m_{\mathrm{R}}^{2} +
\frac{\lambda_{\varphi}}{2} \Phi^{2}}{m_{\mathrm{R}}^{2}} \right) \\ &\quad
+ \frac{m_{\mathrm{R}}^{4} g \bar{\Psi} \Psi}{32 \pi^{2} \left(
m_{\mathrm{R}}^{2} + \frac{\lambda_{\varphi}}{2} \Phi^{2} \right)} + \frac{g
\lambda_{\varphi} \Phi^{2} \bar{\Psi} \Psi}{32 \pi^{2}} \ln \left(
\frac{m_{\mathrm{R}}^{2} + \frac{\lambda_{\varphi}}{2}
\Phi^{2}}{m_{\mathrm{R}}^{2}} \right) \\ &\quad + \frac{g m_{\mathrm{R}}^{2}
\bar{\Psi} \Psi}{16 \pi^{2}} \left[ \ln \left( \frac{m_{\mathrm{R}}^{2} +
\frac{\lambda_{\varphi}}{2} \Phi^{2}}{m_{\mathrm{R}}^{2}} \right) -
\frac{1}{2} \right] \\ &\quad + \frac{M_{\mathrm{E}}^{2}}{\pi^{2} L^{2}}
\sum_{n=1}^{\infty} n^{-2} K_{2}(2 n M_{\mathrm{E}} L) - \frac{M_{g}^{2}}{8
\pi^{2} L^{2}} \sum_{j=1}^{\infty} j^{-2} K_{2}(2 j M_{g} L). \end{aligned}
\label{vr01}
\end{equation}

Having established the explicit form of the renormalized effective potential
for the system of an Elko field interacting with a real scalar field, we can
now proceed to analyze the vacuum energy, its quantum corrections, and the
topological mass of the theory.


\section{Vacuum Energy per Unit Area of the Plates, First-Order
Coupling-Constant Corrections, and Topological Mass}

\label{sec4} 

Knowing the explicit form of the renormalized effective potential presented
in Eq.~(\ref{vr01}), we calculate the vacuum energy per unit area, $A$, of
the plates by evaluating it at the vacuum state given in Eq.~\eqref{vacuum}.
Therefore, we have 
\begin{eqnarray}
\frac{E}{A} &=&LV_{\mathrm{eff}}^{\mathrm{ren}}(v)  \notag \\
&=&\frac{m_{\mathrm{E}}^{2}}{\pi ^{2}L}\sum_{n=1}^{\infty }n^{-2}K_{2}\left(
2nm_{\mathrm{E}}L\right) -\frac{m_{\mathrm{R}}^{2}}{8\pi ^{2}L}%
\sum_{j=1}^{\infty }j^{-2}K_{2}\left( 2jm_{\mathrm{R}}L\right) \,.
\label{ce01}
\end{eqnarray}%
The first term on the r.h.s. of Eq.~\eqref{ce01} corresponds to the Elko
field satisfying Dirichlet boundary conditions, while the second term
represents the contribution from the real scalar field, in agreement with
previous results found in the literature \cite{cruz2020casimir,
Junior:2025thl}. Note that the Elko term is positive, reflecting its
fermionic nature. In contrast with the results found in Refs.~\cite%
{maluf2020casimir, Pereira:2016muy}, the Elko contribution in Eq.~%
\eqref{ce01} is four times larger. This result is further confirmed using
canonical quantization, as shown in Appendix~\ref{cq}. 
\begin{figure}[tbp]
\centering
\includegraphics[width=0.6\textwidth]{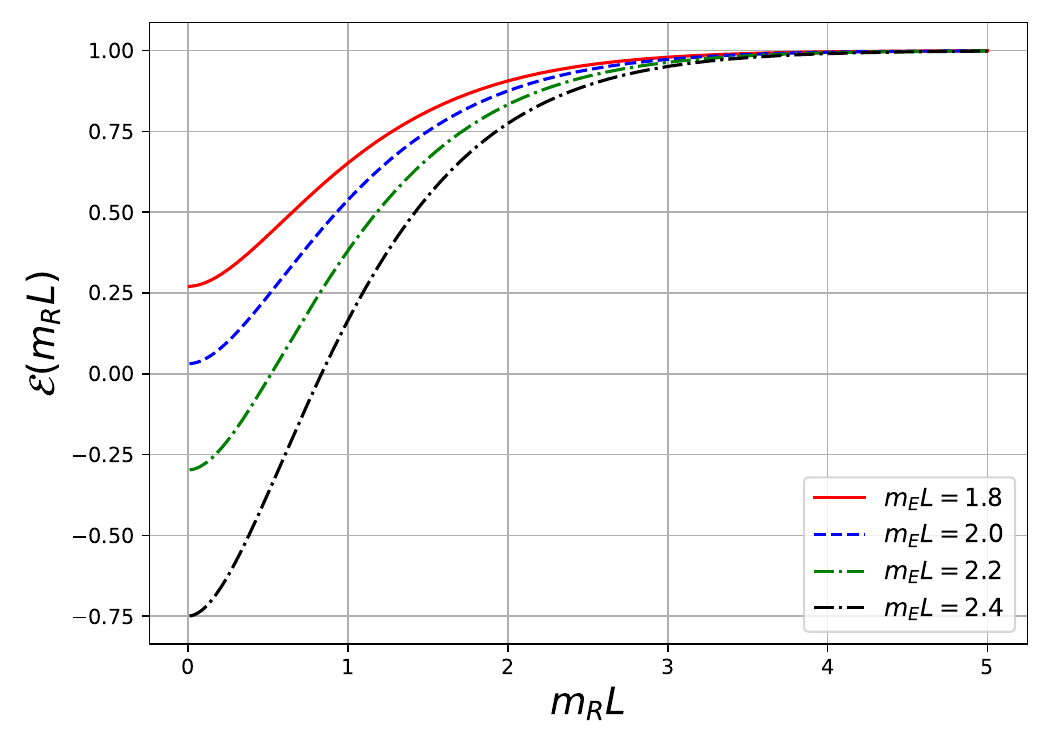}
\caption{Graph of the dimensionless energy $\mathcal{E}\left( m_{\mathrm{R}%
}L\right) $, Eq.~(\protect\ref{de0.1}), as a function of $m_{\mathrm{R}}L$
and fixed values of $m_{\mathrm{E}}L$.}
\label{fig1.1}
\end{figure}

In Figure~\ref{fig1.1} we exhibit  the behavior of dimensionless energy $%
\mathcal{E}\left( m_{\mathrm{R}}L\right) $, as a function of $m_{\mathrm{R}}L
$, defined as%
\begin{equation}
\mathcal{E}\left( m_{\mathrm{R}}L\right) =\frac{EL^{3}\pi ^{2}}{A\varepsilon
_{\mathrm{E}}}=1-\frac{L^{2}m_{\mathrm{R}}^{2}}{8\varepsilon _{\mathrm{E}}}\sum_{j=1}^{\infty
}j^{-2}K_{2}\left( 2jm_{\mathrm{R}}L\right) ,
\label{de0.1}
\end{equation}%
where 
\begin{eqnarray}
\varepsilon _{\mathrm{E}}=\frac{1}{L^{2}m_{\mathrm{E%
}}^{2}}\sum_{n=1}^{\infty }n^{-2}K_{2}\left( 2nm_{\mathrm{E}}L\right). 
\end{eqnarray}
The graph shows that in the
limit of large $m_{\mathrm{R}}L$ the dimensionless energy goes to unity,
that is, the contribution in this case comes only from the Elko field, as it
should be. In addition, increasing the value of $m_{\mathrm{E}}L$ we see
that the vacuum energy decreases.

We may also, in principle, consider the massless limit only for the real scalar field. 
The fundamental anticommutation relations involving the Elko field operators depend on the inverse mass and are therefore ill-defined in the massless limit. 
This behavior is a consequence of the nonlocal nature of the Elko field (see Ref.~\cite{Ahluwalia:2004ab}). 
An additional noteworthy feature of this field is that it becomes local as its mass increases.

The massless limit of the
real field has been investigated in several other works, for instance in
Refs.~\cite{bordag2009advances, cruz2020casimir}. This limit can be obtained
by considering, in Eq.~(\ref{ce01}), 
\begin{equation}
\lim_{z\to 0} z^{\nu} K_{\nu}(z) = 2^{\nu-1} \Gamma(\nu),  \label{limit}
\end{equation}
for the Macdonald function $K_{\nu}(z)$, along with the known value of the
Riemann zeta function $\zeta(4) = \frac{\pi^{4}}{90}$~\cite%
{elizalde1995zeta, Elizalde:1995hck}.


\subsection{First-Order Coupling-Constant Corrections to the Vacuum Energy}

\label{secIVA} 

Using the generalized zeta function and Feynman diagrams, one can compute
the two-loop correction to the vacuum energy per unit area of the plates in
Eq.~\eqref{ce01}, evaluated at the vacuum state $v$. The procedure is
described in Refs.~\cite{Aj,porfirio2021ground,PhysRevD.107.125019,
doi:10.1142/S021827182450069X,toms1980interacting}. The diagrams
corresponding to the self-interaction and cross-interaction contributions
are shown below: 
\begin{eqnarray}
V^{(2)}(v) = \underbrace{ \begin{tikzpicture}[baseline=-0.5ex]
\draw[thick,thick] (0,0) circle (0.4); \draw[thick,thick] (0.8,0) circle
(0.4); \fill (0.4,0) circle (2pt); \node at (-0.3,0.5) {\small \(\varphi\)};
\node at (1.0,0.55) {\small \(\varphi\)}; \end{tikzpicture}}_{\varphi\text{--%
}\varphi \text{ loop}} \quad+ \quad \underbrace{ %
\begin{tikzpicture}[baseline=-0.5ex] \draw[thick] (0,0) circle (0.4);
\draw[thick,dashed] (0.8,0) circle (0.4); \fill (0.4,0) circle (2pt); \node
at (-0.4,0.5) {\small \(\eta\)}; \node at (1.2,0.5) {\small \(\varphi\)};
\end{tikzpicture}}_{\eta\text{--}\varphi \text{ mixed loop}} \quad
\label{Fdiagrams}
\end{eqnarray}

The contribution from the real scalar field, due to the self-interaction
term $\frac{\lambda_{\varphi}}{4!}\varphi^4$, can be obtained from the first
diagram on the r.h.s. of Eq.~\eqref{Fdiagrams}. Mathematically, this
translates into 
\begin{align}
V_{\lambda_{\varphi}}^{(2)}(0) &= \frac{3\lambda_{\varphi}}{4!} \left. \left[
\frac{\zeta_{\varphi}^{\mathrm{ren}}(1)}{\Omega_4} \right]^2 \right|_{v} 
\notag \\
&= \frac{\lambda_{\varphi} m_{\mathrm{R}}^2}{128 \pi^4 L^2} \left[
\sum_{n=1}^{\infty} \frac{1}{n} K_1(2 n m_{\mathrm{R}} L) \right]^2 \,,
\label{v2h}
\end{align}
where $\zeta_{\varphi}^{\mathrm{ren}}(1)$ is the generalized zeta function
defined in Eq.~(\ref{zh01}), with both the divergent part at $s=1$ (the
Minkowski contribution) and the contribution from a single plate subtracted 
\cite{Aj,porfirio2021ground,PhysRevD.107.125019,
doi:10.1142/S021827182450069X}, namely, 
\begin{equation}
\zeta_{\varphi}^{\mathrm{ren}}(s) = \zeta_{\varphi}(s) - \frac{\Omega_4
M_g^{4-2s}}{16 \pi^2} \frac{\Gamma(s-2)}{\Gamma(s)} + \frac{\Omega_4
M_g^{3-2s}}{16 \pi^{\frac{3}{2}} L} \frac{\Gamma(s-3/2)}{\Gamma(s)} \,.
\label{renZeta}
\end{equation}

Note that the factor of three in Eq.~\eqref{v2h} originates from the Wick's theorem for field contractions, while the symmetry factor in this case is $%
\tfrac{1}{2}$. The corresponding loop implies that the zeta function in Eq.~%
\eqref{renZeta} should appear divided by $\Omega_4$. Moreover, for the
diagram under consideration there is a single vertex contributing with a
factor $\tfrac{\lambda_{\varphi}}{4!}$.

Similarly, the first-order correction arising from the interaction between
the fields is proportional to the coupling constant $g$, i.e., 
\begin{align}
V_g^{(2)}(0) &= g \left. \left[ \frac{\zeta_{\eta}^{\mathrm{ren}}(1)}{%
\Omega_4} \, \frac{\zeta_{\varphi}^{\mathrm{ren}}(1)}{\Omega_4} \right]
\right|_{v}  \notag \\
&= g \frac{m_{\mathrm{E}} m_{\mathrm{R}}}{16 \pi^4 L^2} \left[
\sum_{n=1}^{\infty} \frac{1}{n} K_1(2 n m_{\mathrm{E}} L) \right] \left[
\sum_{j=1}^{\infty} \frac{1}{j} K_1(2 j m_{\mathrm{R}} L) \right] \,,
\label{v2i}
\end{align}
where $\zeta_{\eta}^{\mathrm{ren}}(1)$ is defined analogously to $%
\zeta_{\varphi}^{\mathrm{ren}}(1)$. Note that the diagram representing the
mixed terms in Eq.~\eqref{Fdiagrams} indicates that the corresponding
expression involves the product of the coupling constant $g$ (from a single
vertex) with the renormalized zeta functions (\ref{zh01}) and (\ref{ze01})
at $s=1$, each divided by $\Omega_4$.

Taking into account the two previous expressions, the total first-order
correction to the vacuum energy per unit area of the plates in Eq.~%
\eqref{ce01} is then given by%
\begin{eqnarray}
\frac{E^{(2)}}{A} &=&\frac{\lambda _{\varphi }m_{\mathrm{R}}^{2}}{128\pi
^{4}L}\left[ \sum_{n=1}^{\infty }\frac{1}{n}K_{1}(2nm_{\mathrm{R}}L)\right]
^{2}  \notag \\
&&+\frac{gm_{\mathrm{E}}m_{\mathrm{R}}}{16\pi ^{4}L}\left[
\sum_{n=1}^{\infty }\frac{1}{n}K_{1}(2nm_{\mathrm{E}}L)\right] \left[
\sum_{j=1}^{\infty }\frac{1}{j}K_{1}(2jm_{\mathrm{R}}L)\right] \,.
\end{eqnarray}

Once again, the massless limit of the Elko field is not well-defined,
whereas in the case of the real field it can be obtained by applying the
limit given in Eq.~\eqref{limit}.

From this point onward, we shall consider the mass corrections to the
fields, which arise from the boundary conditions and the interactions in the
theory. The topological mass is computed from the renormalization conditions
in Eq.~(\ref{rc14.5.2}); however, we must take into account the renormalized
effective potential. We first analyze the topological mass associated with
the Elko field, and then the mass correction corresponding to the Higgs
field.


\subsection{Topological mass of the Elko field}

\label{secIVB} 
We first consider the mass correction to the Elko field. For this purpose,
we have to compute two derivatives of the renormalized effective potential
at the vacuum state, Eq.~(\ref{rc14.5.2}), namely, 
\begin{equation}
m_{\mathrm{T}}^{2}=\left. \frac{d^{2}V_{\text{\textrm{eff}}}^{\mathrm{ren}}}{%
d\Psi\, d\bar{\Psi}} \right\vert _{v}\,.
\end{equation}
In the effective potential given in Eq.~(\ref{vr01}), the first term in the
last line does not depend on $\bar{\Psi}$ or $\Psi$. Therefore, we focus on
the second term, since all the other terms vanish once the vacuum state is
taken. Because $\bar{\Psi}$ and $\Psi$ are Grassmann variables, the
derivatives cannot be taken directly. First, we expand the term $M_{g}$ in
the argument of the Macdonald function $K_{\gamma}(z)$ in Eq.~(\ref{vr01})
with respect to $\Psi\bar{\Psi}$ up to first order, using the properties of
Grassmann variables. This yields 
\begin{equation}
M_{g}=\left( m_{\mathrm{H}}^{2}+\frac{\lambda_{\varphi}}{2}\Phi^{2} +2g\,%
\bar{\Psi}\Psi \right)^{\frac{1}{2}} = M_{\Phi}+\frac{g\,\bar{\Psi}\Psi}{%
M_{\Phi}}\,,  \label{expansion1}
\end{equation}
where we have defined the quantity $M_{\Phi}$ as 
\begin{equation}
M_{\Phi}=\left( m_{\mathrm{H}}^{2}+\frac{\lambda_{\varphi}}{2}%
\Phi^{2}\right)^{\frac{1}{2}}\,.
\end{equation}
Note that, due to the Grassmann nature of the variables $\Psi\bar{\Psi}$, which are proportional to a single Grassmann variable, all higher-order terms in Eq.~\eqref{expansion1} vanish.
\begin{figure}[!h]
\includegraphics[width=.5\textwidth]{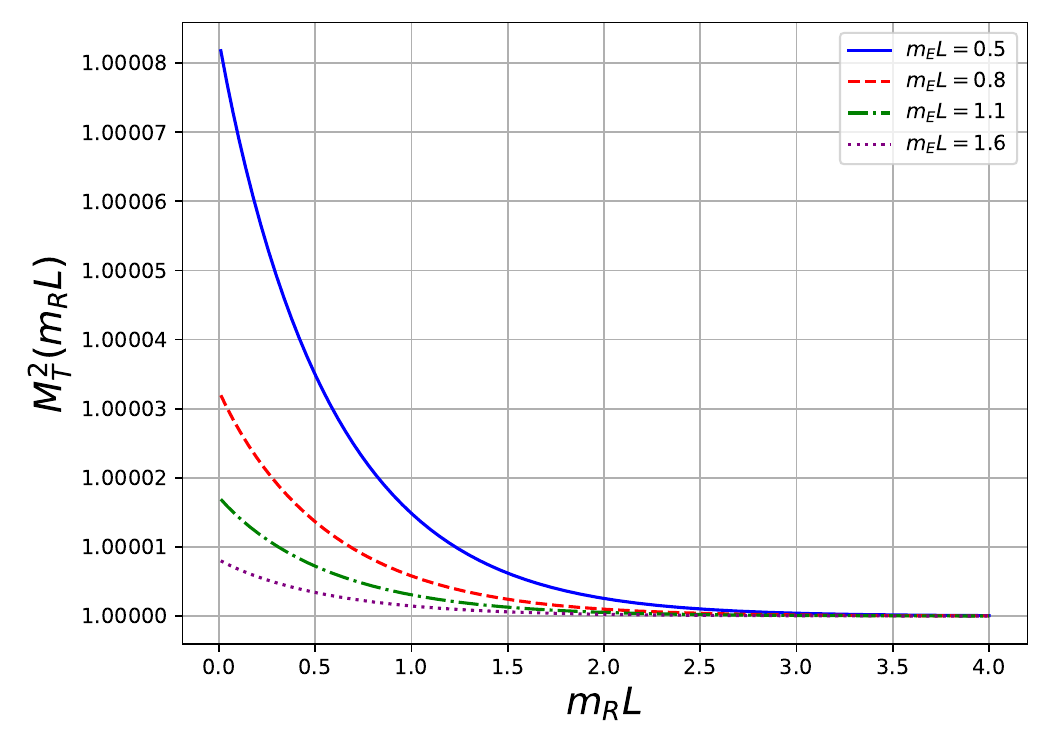}\hfill
\caption{Graph of $M_{\mathrm{T}}^{2}\left( m_{\mathrm{R}}L\right) $, Eq.~(%
\protect\ref{dtm0.1}), as a function of $m_{\mathrm{R}}L$, with $g=10^{-3}$. 
}
\label{fig1.2}
\end{figure}

The next step is to use the following integral representation of the
Macdonald function~\cite{gradshteyn2014table}: 
\begin{equation}
K_{\alpha }(x)=\int_{0}^{\infty }dt\,\cosh (\alpha t)\,e^{-x\cosh t}\,,
\end{equation}%
which leads to 
\begin{equation}
K_{2}\!\left( jM_{g}L\right) =\int_{0}^{\infty }dt\,\cosh (2t)\,\exp \!\left[
-jL\!\left( M_{\Phi }+\frac{g\,\bar{\Psi}\Psi }{M_{\Phi }}\right) \cosh t%
\right] .
\end{equation}%
Expanding the exponential in the above equation and using the properties of
the Grassmann variables, we obtain the Macdonald function in the following
form: 
\begin{equation}
K_{2}\!\left( jM_{g}L\right) =K_{2}\!\left( jM_{\Phi }L\right) +jgL\,\frac{%
\bar{\Psi}\Psi }{M_{\Phi }}\,\frac{dK_{2}\!\left( jM_{\Phi }L\right) }{%
d\!\left( jM_{\Phi }L\right) }\,.  \label{k2}
\end{equation}%
In this form, it is straightforward to take the derivatives. Therefore, we
obtain the topological mass of the Elko field as 
\begin{equation}
m_{\mathrm{T}}^{2}=m_{\mathrm{E}}^{2}+\frac{g\,m_{\mathrm{R}}}{4\pi ^{2}L}%
\sum_{j=1}^{\infty }j^{-1}K_{1}\!\left( 2j\,m_{\mathrm{R}}L\right) .
\label{me0.1}
\end{equation}%
Note that the field acquires a mass correction proportional to the coupling
constant $g$ and exhibits a dependence on the distance $L$ (related to the
boundary conditions) as well as on the mass of the scalar field $m_{\mathrm{R%
}}$.

In Figure~\ref{fig1.2} we we exhibit the behavior of dimensionless
topological mass squared associeted with the Elko field,%
\begin{equation}
M_{\mathrm{T}}^{2}\left( m_{\mathrm{R}}L\right) =\frac{m_{\mathrm{T}}^{2}}{%
m_{\mathrm{E}}^{2}}=1+\frac{g\,m_{\mathrm{R}}L}{4\pi ^{2}m_{\mathrm{E}%
}^{2}L^{2}}\sum_{j=1}^{\infty }j^{-1}K_{1}\!\left( 2j\,m_{\mathrm{R}%
}L\right) \,,  \label{dtm0.1}
\end{equation}%
as a function of $m_{\mathrm{R}}L$, with fixed value $g=10^{-3}$ and $m_{%
\mathrm{E}}L=0.5$, $0.8$, $1.1$, $1.6$. As already mentioned, the massless limit of the above expression can be
taken only for the real scalar field, using Eq.~\eqref{limit}.

Next, we consider the topological mass associated with the real scalar field.


\subsection{Topological Mass of the Real Scalar Field}
\label{secIVC} 
%
Considering the real scalar field, the derivatives of the renormalized
effective potential is expressed as 
\begin{equation*}
m_{\mathrm{T}}^{2}=\left. \frac{d^{2}V_{\mathrm{eff}}^{\mathrm{ren}}}{d\Phi
^{2}}\right\vert _{v}.
\end{equation*}%
This provides the following expression for the topological mass of the real
scalar field 
\begin{eqnarray}
m_{\mathrm{T}}^{2} &=&m_{\mathrm{R}}^{2}-\frac{2gm_{\mathrm{E}}}{\pi ^{2}L}%
\sum_{n=1}^{\infty }n^{-1}K_{1}\left( 2nm_{\mathrm{E}}L\right)  \notag \\
&&+\frac{\lambda _{\varphi }m_{\mathrm{R}}}{8\pi ^{2}L}\sum_{j=1}^{\infty
}j^{-1}K_{1}\left( 2jm_{\mathrm{R}}L\right) \,.
\end{eqnarray}%
\begin{figure}[!h]
\includegraphics[width=.5\textwidth]{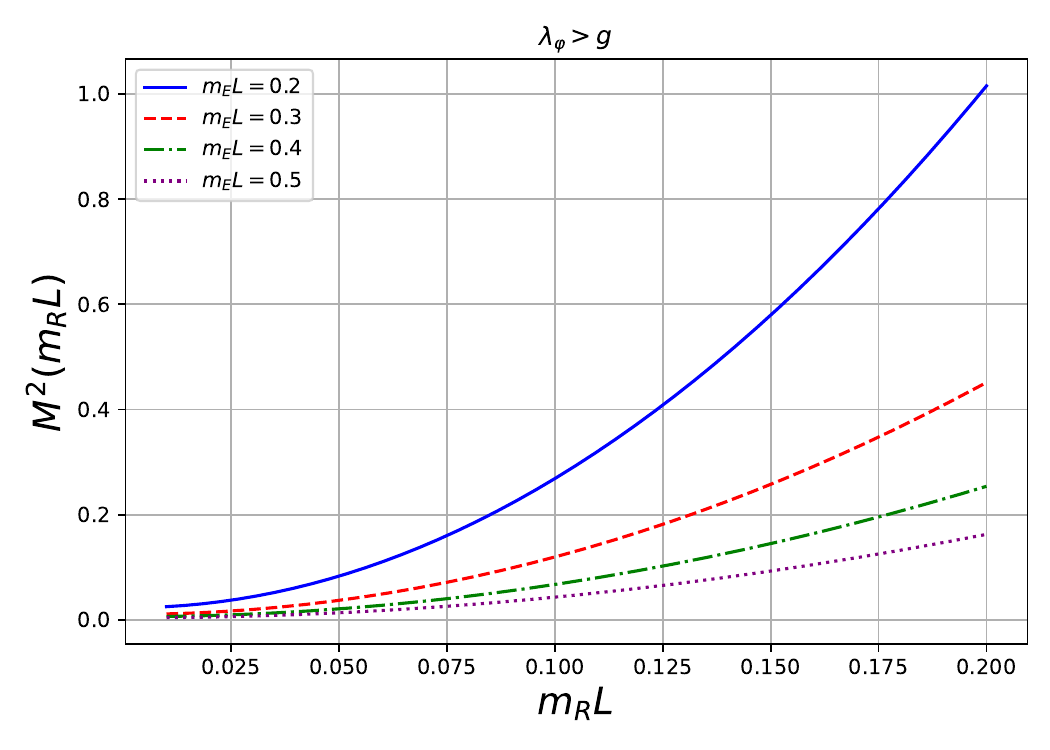}\hfill %
\includegraphics[width=.5\textwidth]{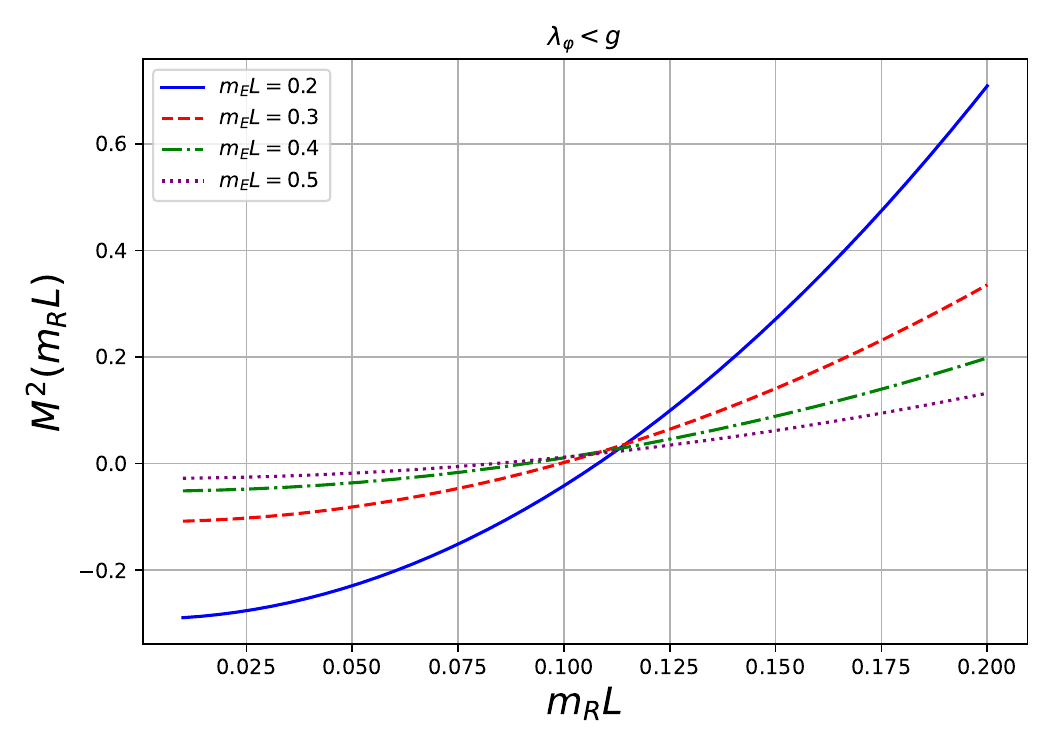}\hfill
\caption{ The graph on the left shows the $M^{2}\left( m_{\mathrm{R}%
}L\right) $, Eq.~(\protect\ref{dtm0.2}), with $g=10^{-3}$ and $\protect%
\lambda _{\protect\varphi }=10^{-1}$, as a function of $m_{\mathrm{R}}L$,
while the graph on the right takes the values $g=10^{-1}$ and $\protect%
\lambda _{\protect\varphi }=10^{-3}$.}
\label{fig1.2a}
\end{figure}
Note that one term of the correction is proportional to the coupling constant 
$g$, since it comes from the interaction between the fields and it also
depends on the mass of the Elko field. The other correction comes from the
self-interaction of the real scalar field, so it is proportional to the
coupling constant $\lambda _{\varphi }$. Moreover, the two terms also
depends on the parameter $L$ related with the boundary condition.

Figure~\ref{fig1.2a} shows the graph of the dimensionless topological mass
associated with the real field,%
\begin{eqnarray}
M^{2}\left( m_{\mathrm{R}}L\right) &=&\frac{m_{\mathrm{T}}^{2}}{m_{\mathrm{E}%
}^{2}}=\frac{m_{\mathrm{R}}^{2}L^{2}}{m_{\mathrm{E}}^{2}L^{2}}-\frac{2g}{\pi
^{2}m_{\mathrm{E}}L}\sum_{n=1}^{\infty }n^{-1}K_{1}\left( 2nm_{\mathrm{E}%
}L\right)  \notag \\
&&+\frac{\lambda _{\varphi }m_{\mathrm{R}}L}{8\pi ^{2}m_{\mathrm{E}}^{2}L^{2}%
}\sum_{j=1}^{\infty }j^{-1}K_{1}\left( 2jm_{\mathrm{R}}L\right) \,,
\label{dtm0.2}
\end{eqnarray}%
as a function of $m_{\mathrm{R}}L$ and fixed values $m_{\mathrm{E}}L=0.2$, $%
0.3$, $0.4$, $0.5$. The graph on the left takes the values $g=10^{-3}$ and $%
\lambda _{\varphi }=10^{-1}$. As we can see, the graph on the left is
strictely positive, since $\lambda _{\varphi }>g$. However, the graph on the
right shows possibilites for negative values of the squared mass, which in
principle indicates vacuum instability \cite{PhysRevD.107.125019}.

\section{Concluding remarks}

\label{sec5} 

In this work, we have analyzed the quantum vacuum properties of a
four-dimensional system composed of a mass-dimension-one fermionic field
(Elko) interacting with a real scalar field through a quadratic coupling,
both subjected to Dirichlet boundary conditions on two parallel plates
separated by a distance L. Using the path integral formalism and the
effective potential method, we evaluated the renormalized vacuum energy
density, including first-order corrections due to the interaction term, and
derived the corresponding topological mass corrections for each field.

Our results show that the presence of boundaries and interactions produce
nontrivial modifications to the vacuum structure of the theory. In
particular, the contribution of the Elko field to the vacuum energy is
significantly enhanced compared to that of the scalar field, while the
first-order coupling corrections generate boundary-induced mass shifts that
depend explicitly on the plate separation. These topological mass
corrections encode the interplay between quantum fluctuations, geometry, and
field interactions, highlighting how non-standard fermionic fields can lead
to distinctive signatures in confined geometries.

In addition to the analytical results, we exhibit numerically the depence of
the vacuum energy associated with the scalar and Elko fields as function of $%
m_{R}L$ in Fig. \ref{fig1.1}. Also we present in Figs. \ref{fig1.2} and \ref%
{fig1.2a}, the behavior of the topological mass associated with the Elko and
scalar fields, respectively. In all these graphs, we observe that they go to
zero in the limit of large value for $L$.

Moreover, the analysis indicates that, depending on the parameter regime,
the scalar sector may develop negative topological mass corrections,
signaling potential boundary-induced instabilities or spontaneous symmetry
breaking. Such effects could point to the existence of multiple vacua or
metastable configurations, a feature that merits further investigation
through a detailed vacuum stability analysis.

Although our study focused on a real scalar field for simplicity, the
formalism can be directly extended to complex scalar fields, including the
Higgs doublet, preserving the essential features of the Elko - scalar
interaction. Future work may explore higher-loop corrections,
finite-temperature effects, or curved-space generalizations, aiming to
connect these boundary-induced quantum phenomena with broader aspects of
Elko phenomenology and dark matter models.

\bigskip

{\acknowledgments} The authors are grateful to Prof.~D.~Grumiller and Prof.~Andrea Erdas for useful discussions. H.F.S.M. is partially supported by the Brazilian agency
National Council for Scientific and Technological Development (CNPq) under
Grant No. 308049/2023-3. The author E.R.B.M. thanks CNPq for partial
support, Grant No. 304332/2024-0.

\bigskip

\appendix


\section{On renormalization}

\label{ren} 

In this section, we apply the renormalization conditions presented in Eq.~(%
\ref{rc14.5}). These conditions are imposed on the effective potential, Eq.~(%
\ref{ve01}), in the Minkowski spacetime limit, that is, $L \rightarrow
\infty $. In addition, terms proportional to the third power of the masses
are neglected, as is customary when considering Dirichlet boundary
conditions~\cite{cruz2020casimir}. Therefore, the part of the effective
potential relevant for the renormalization procedure reads 
\begin{eqnarray}
V_{\text{eff}}&=& m_{\mathrm{E}}^{2}\bar{\Psi}\Psi +\frac{1}{2}m_{\mathrm{H}%
}^{2}\Phi^{2} +\frac{\lambda_{\varphi}}{4!}\Phi^{4} +g\Phi^{2}\bar{\Psi}\Psi
\notag \\
&&+\frac{C_{1}}{4!}\Phi^{4} +\frac{C_{2}}{2}\Phi^{2} +C_{3} +C_{4}\bar{\Psi}%
\Psi +C_{5}\Phi^{2}\bar{\Psi}\Psi  \notag \\
&&+\frac{M_{\mathrm{E}}^{4}}{2^{3}\pi^{2}} \left( \frac{3}{2} - \ln \frac{M_{%
\mathrm{E}}^{2}}{\alpha^{2}} \right) +\frac{M_{g}^{4}}{2^{6}\pi^{2}} \left(
\ln \frac{M_{g}^{2}}{\beta^{2}} - \frac{3}{2} \right) .  \label{re0.1}
\end{eqnarray}

We first apply the renormalization conditions associated with the scalar
field. The first condition in Eq.~(\ref{rc14.5}) gives 
\begin{equation}
C_{1} = \frac{3}{\pi^{2}} g^{2} \ln \frac{m_{\mathrm{E}}^{2}}{\alpha^{2}} - 
\frac{3\lambda_{\varphi}^{2}}{2^{5}\pi^{2}} \ln \frac{m_{\mathrm{H}}^{2}}{%
\beta^{2}} .  \label{re0.2}
\end{equation}
The renormalization condition~\eqref{rc14.5.2} associated with the scalar
field determines the constant $C_{2}$ as 
\begin{equation}
C_{2} = \frac{\lambda_{\varphi} m_{\mathrm{H}}^{2}}{2^{5}\pi^{2}} \left( 1 -
\ln \frac{m_{\mathrm{H}}^{2}}{\beta^{2}} \right) + \frac{g m_{\mathrm{E}}^{2}%
}{2\pi^{2}} \left( \ln \frac{m_{\mathrm{E}}^{2}}{\alpha^{2}} - 1 \right) .
\label{re0.3}
\end{equation}
Next, applying the third condition in Eq.~(\ref{C3}), we obtain the constant 
$C_{3}$ as 
\begin{equation}
C_{3} = -\frac{m_{\mathrm{E}}^{4}}{2^{3}\pi^{2}} \left( \frac{3}{2} - \ln 
\frac{m_{\mathrm{E}}^{2}}{\alpha^{2}} \right) -\frac{m_{\mathrm{H}}^{4}}{%
2^{6}\pi^{2}} \left( \ln \frac{m_{\mathrm{H}}^{2}}{\beta^{2}} - \frac{3}{2}
\right) .  \label{re0.4}
\end{equation}

After substituting the renormalization constants~(\ref{re0.2})-(\ref{re0.4}) into Eq.~(\ref{re0.1}), the renormalized effective potential
becomes 
\begin{eqnarray}
V_{\text{eff}} &=& m_{\mathrm{E}}^{2}\bar{\Psi}\Psi +\frac{1}{2}m_{\mathrm{H}%
}^{2}\Phi^{2} +\frac{\lambda_{\varphi}}{4!}\Phi^{4} +g\Phi^{2}\bar{\Psi}\Psi
+\xi(\bar{\Psi}, \Psi, C_{i})  \notag \\
&&+\frac{g^{2}\Phi^{4}}{2^{3}\pi^{2}} \left[ \frac{3}{2} - \ln \left( \frac{%
M_{\mathrm{E}}^{2}}{m_{\mathrm{E}}^{2}} \right) \right] +\frac{%
\lambda_{\varphi}^{2}\Phi^{4}}{2^{8}\pi^{2}} \left[ \ln \left( \frac{%
M_{g}^{2}}{m_{\mathrm{H}}^{2}} \right) - \frac{3}{2} \right]  \notag \\
&&+\frac{g m_{\mathrm{E}}^{2}\Phi^{2}}{2^{2}\pi^{2}} \left[ \frac{1}{2} -
\ln \left( \frac{M_{\mathrm{E}}^{2}}{m_{\mathrm{E}}^{2}} \right) \right] +%
\frac{\lambda_{\varphi} m_{\mathrm{H}}^{2}\Phi^{2}}{2^{6}\pi^{2}} \left[ \ln
\left( \frac{M_{g}^{2}}{m_{\mathrm{H}}^{2}} \right) - \frac{1}{2} \right] 
\notag \\
&&-\frac{m_{\mathrm{E}}^{4}}{2^{3}\pi^{2}} \ln \left( \frac{M_{\mathrm{E}%
}^{2}}{m_{\mathrm{E}}^{2}} \right) +\frac{m_{\mathrm{H}}^{4}}{2^{6}\pi^{2}}
\ln \left( \frac{M_{g}^{2}}{m_{\mathrm{H}}^{2}} \right)  \notag \\
&&+\frac{(4m_{\mathrm{H}}^{2} g \bar{\Psi}\Psi +2\lambda_{\varphi} g
\Phi^{2} \bar{\Psi}\Psi)}{2^{6}\pi^{2}} \left[ \ln \left( \frac{M_{g}^{2}}{%
\beta^{2}} \right) - \frac{3}{2} \right],  \label{re0.5}
\end{eqnarray}
where we take $(\bar{\Psi}\Psi)^2 = 0$, since it is proportional to a single Grassmann variable, and
\begin{equation}
\xi(\bar{\Psi}, \Psi, C_{i}) = C_{4}\bar{\Psi}\Psi + C_{5}\Phi^{2}\bar{\Psi}%
\Psi \,.  \label{re0.6}
\end{equation}

Now we take the renormalization conditions associated with the Elko field.
First, we rewrite the logarithmic function as 
\begin{equation}
\ln \left( \frac{M_{g}^{2}}{\beta^{2}} \right) = \ln \left( \frac{m_{\mathrm{%
H}}^{2} + \frac{\lambda_{\varphi}}{2}\Phi^{2}}{\beta^{2}} \right) + \ln
\left( 1 + \frac{2g\bar{\Psi}\Psi}{m_{\mathrm{H}}^{2} + \frac{%
\lambda_{\varphi}}{2}\Phi^{2}} \right),  \label{re0.7}
\end{equation}
and, using the expansion 
\begin{equation}
\ln(1 + x) = \sum_{n=1}^{\infty}(-1)^{n+1}\frac{x^{n}}{n},  \label{re0.8}
\end{equation}
together with the fermionic property of the Elko field, we find 
\begin{equation}
\ln \left( \frac{M_{g}^{2}}{\beta^{2}} \right) = \ln \left( \frac{m_{\mathrm{%
H}}^{2} + \frac{\lambda_{\varphi}}{2}\Phi^{2}}{\beta^{2}} \right) + \frac{2g%
\bar{\Psi}\Psi}{m_{\mathrm{H}}^{2} + \frac{\lambda_{\varphi}}{2}\Phi^{2}}.
\label{re0.9}
\end{equation}

We can now apply the condition associated with the mass of the Elko field,
that is, the second condition in Eq.~(\ref{rc14.5.2}), which gives the
fourth renormalization constant as 
\begin{equation}
C_{4} = -\frac{g m_{\mathrm{H}}^{2}}{2^{5}\pi^{2}} - \frac{4 m_{\mathrm{H}%
}^{2} g}{2^{6}\pi^{2}} \left( \ln \frac{m_{\mathrm{H}}^{2}}{\beta^{2}} - 
\frac{3}{2} \right).  \label{re1.0}
\end{equation}
Applying the second condition in Eq.~(\ref{rc14.5}), we obtain the last
renormalization constant: 
\begin{equation}
C_{5} = -\frac{g \lambda_{\varphi}}{2^{5}\pi^{2}} \left( \ln \frac{m_{%
\mathrm{H}}^{2}}{\beta^{2}} - \frac{3}{2} \right).  \label{re1.1}
\end{equation}

Substituting the renormalization constants obtained in Eqs.~(\ref{re1.0})
and~(\ref{re1.1}) into the potential of Eq.~(\ref{re0.5}), and adding the
terms that depend on $L$, we obtain the renormalized effective potential
associated with the system studied here, that is, 
\begin{eqnarray}
V_{\text{eff}}^{\mathrm{ren}}(\bar{\Psi}, \Psi, \Phi) &=& m_{\mathrm{E}}^{2}%
\bar{\Psi}\Psi + \frac{1}{2}m_{\mathrm{H}}^{2}\Phi^{2} + \frac{%
\lambda_{\varphi}}{4!}\Phi^{4} + g\Phi^{2}\bar{\Psi}\Psi  \notag \\
&& + \frac{g^{2}\Phi^{4}}{2^{3}\pi^{2}} \left[ \frac{3}{2} - \ln\left( \frac{%
M_{\mathrm{E}}^{2}}{m_{\mathrm{E}}^{2}} \right) \right] + \frac{%
\lambda_{\varphi}^{2}\Phi^{4}}{2^{8}\pi^{2}} \left[ \ln\left( \frac{M_{g}^{2}%
}{m_{\mathrm{H}}^{2}} \right) - \frac{3}{2} \right]  \notag \\
&& + \frac{g m_{\mathrm{E}}^{2}\Phi^{2}}{2^{2}\pi^{2}} \left[ \frac{1}{2} -
\ln\left( \frac{M_{\mathrm{E}}^{2}}{m_{\mathrm{E}}^{2}} \right) \right] + 
\frac{\lambda_{\varphi} m_{\mathrm{H}}^{2}\Phi^{2}}{2^{6}\pi^{2}} \left[
\ln\left( \frac{M_{g}^{2}}{m_{\mathrm{H}}^{2}} \right) - \frac{1}{2} \right]
\notag \\
&& - \frac{m_{\mathrm{E}}^{4}}{2^{3}\pi^{2}} \ln\left( \frac{M_{\mathrm{E}%
}^{2}}{m_{\mathrm{E}}^{2}} \right) + \frac{m_{\mathrm{H}}^{4}}{2^{6}\pi^{2}}
\ln\left( \frac{m_{\mathrm{H}}^{2} + \frac{\lambda_{\varphi}}{2}\Phi^{2}}{m_{%
\mathrm{H}}^{2}} \right)  \notag \\
&& + \frac{m_{\mathrm{H}}^{4} g \bar{\Psi}\Psi}{2^{5}\pi^{2} \left( m_{%
\mathrm{H}}^{2} + \frac{\lambda_{\varphi}}{2}\Phi^{2} \right)} + \frac{g
\lambda_{\varphi} \Phi^{2}\bar{\Psi}\Psi}{2^{5}\pi^{2}} \ln\left( \frac{m_{%
\mathrm{H}}^{2} + \frac{\lambda_{\varphi}}{2}\Phi^{2}}{m_{\mathrm{H}}^{2}}
\right)  \notag \\
&& + \frac{g m_{\mathrm{H}}^{2}\bar{\Psi}\Psi}{2^{4}\pi^{2}} \left[
\ln\left( \frac{m_{\mathrm{H}}^{2} + \frac{\lambda_{\varphi}}{2}\Phi^{2}}{m_{%
\mathrm{H}}^{2}} \right) - \frac{1}{2} \right]  \notag \\
&& + \frac{M_{\mathrm{E}}^{2}}{\pi^{2}L^{2}} \sum_{n=1}^{\infty} n^{-2}
K_{2}(2n M_{\mathrm{E}} L) - \frac{M_{g}^{2}}{2^{3}\pi^{2}L^{2}}
\sum_{j=1}^{\infty} j^{-2} K_{2}(2j M_{g} L).
\end{eqnarray}


\section{Canonical quantization}

\label{cq} 
In this appendix we compute the vacuum energy using the energy--momentum
tensor. The field expansion for the Elko field reads \cite{Pereira:2016muy} 
\begin{eqnarray}
\hat{\eta}(x) &=& \int \frac{d^{3}p}{(2\pi)^{3/2}\sqrt{2mE}} \sum_{\alpha} %
\left[ \hat{a}_{\alpha}(\mathbf{p})\, \lambda_{\alpha}^{S}(\mathbf{p}%
)\,e^{-ip_{\mu}x^{\mu}} + \hat{a}_{\alpha}^{\dagger}(\mathbf{p})\,
\lambda_{\alpha}^{A}(\mathbf{p})\,e^{ip_{\mu}x^{\mu}} \right],  \notag \\
\hat{\bar{\eta}}(x) &=& \int \frac{d^{3}p}{(2\pi)^{3/2}\sqrt{2mE}}
\sum_{\alpha^{\prime }} \left[ \hat{a}_{\alpha}^{\dagger}(\mathbf{p})\, \bar{%
\lambda}_{\alpha}^{S}(\mathbf{p})\,e^{ip_{\mu}x^{\mu}} + \hat{a}_{\alpha}(%
\mathbf{p})\, \bar{\lambda}_{\alpha}^{A}(\mathbf{p})\,e^{-ip_{\mu}x^{\mu}} %
\right].  \label{fe1}
\end{eqnarray}
The index $\alpha$ runs over four possibilities, corresponding to the two
self-conjugate ($S$) and two anti-self-conjugate ($A$) spinors. These obey $%
\hat{C}\lambda_{\alpha}^{S/A}(\mathbf{p}) = \pm \lambda_{\alpha}^{S/A}(%
\mathbf{p})$, where $\hat{C}$ is the charge conjugation operator. The
creation and annihilation operators satisfy the anticommutation algebra 
\begin{equation}
\bigl[ \hat{a}_{\alpha}(\mathbf{p}),\, \hat{a}_{\alpha^{\prime }}^{\dagger}(%
\mathbf{p}^{\prime }) \bigr]_{+} = \delta_{\alpha \alpha^{\prime }}\,\delta(%
\mathbf{p}-\mathbf{p}^{\prime }),  \label{at1}
\end{equation}
with all other anticommutators vanishing.

The free Lagrangian for the Elko field is 
\begin{equation}
\mathcal{L} = \partial^{\mu}\bar{\eta}\,\partial_{\mu}\eta - m_{E}^{2}\,\bar{%
\eta}\eta .  \label{le1}
\end{equation}
Using this Lagrangian together with Eqs.~(\ref{fe1}) and (\ref{at1}), and
the spinor identities \cite{Pereira:2016muy,ahluwalia2017theory} 
\begin{eqnarray}
\bar{\lambda}_{\alpha^{\prime }}^{S}(\mathbf{p})\lambda_{\alpha}^{S}(\mathbf{%
p}) &=& 2m_{E}\,\delta_{\alpha^{\prime }\alpha},  \notag \\
\bar{\lambda}_{\alpha^{\prime }}^{A}(\mathbf{p})\lambda_{\alpha}^{A}(\mathbf{%
p}) &=& -2m_{E}\,\delta_{\alpha^{\prime }\alpha},  \notag \\
\bar{\lambda}_{\alpha^{\prime }}^{S}(\mathbf{p})\lambda_{\alpha}^{A}(\mathbf{%
p}) & = & \bar{\lambda}_{\alpha^{\prime }}^{A}(\mathbf{p})\lambda_{%
\alpha}^{S}(\mathbf{p}) =0,
\end{eqnarray}
we compute the vacuum energy using 
\begin{equation}
E_{0}=\int d^{3}x\,\langle 0 |\, T^{0}_{\ 0}\, |0\rangle ,
\end{equation}
where $|0\rangle$ is the Fock vacuum, $\hat{a}_{\alpha}(\mathbf{p}%
)|0\rangle=0$, and $T^{0}_{\ 0}$ is the energy--momentum tensor, 
\begin{equation}
T^{0}_{\ 0} = \frac{\partial\mathcal{L}}{\partial(\partial_{0}\eta)}%
\partial_{0}\eta + \partial_{0}\bar{\eta}\, \frac{\partial\mathcal{L}}{%
\partial(\partial_{0}\bar{\eta})} - \mathcal{L}.
\end{equation}

A direct computation yields 
\begin{equation}
\int d^{3}x\, T^{0}_{\ 0} = 2\int d^{3}p\,\omega\sum_{\alpha} \left[ \hat{a}%
_{\alpha}^{\dagger}(\mathbf{p})\,\hat{a}_{\alpha}(\mathbf{p}) - \hat{a}%
_{\alpha}(\mathbf{p})\,\hat{a}_{\alpha}^{\dagger}(\mathbf{p}) \right].
\end{equation}
Using the anticommutation relation (\ref{at1}) we obtain, under Dirichlet
boundary conditions, 
\begin{equation}
E_{0} = -2\sum_{n=1}^{\infty}\! \int d^{2}p_{\parallel}\,
\omega\sum_{\alpha}\, \delta(\mathbf{p}_{\parallel}-\mathbf{p}_{\parallel})
= -4\sum_{n=1}^{\infty}\! \int d^{2}p_{\parallel}\,\omega\,\delta(0),
\label{vee1}
\end{equation}
where the subscript $\parallel$ denotes momenta parallel to the plates, and 
\begin{equation}
\omega=\sqrt{\frac{n^{2}\pi^2}{L^2}+ \mathbf{p}_{\parallel}^{2} + m^{2}}.
\end{equation}

The distribution $\delta(0)$ is evaluated using its Fourier representation, 
\begin{equation}
\delta(\mathbf{p}_{\parallel}-\mathbf{p}^{\prime }_{\parallel}) = \frac{1}{%
(2\pi)^{2}} \int e^{i\mathbf{r}_{\parallel}\cdot (\mathbf{p}_{\parallel}-%
\mathbf{p}^{\prime }_{\parallel})}\, d^{2}r_{\parallel} ,
\end{equation}
yielding, for $\mathbf{p}_{\parallel}=\mathbf{p}^{\prime }_{\parallel}$, 
\begin{equation}
\delta(0)=\frac{A}{(2\pi)^{2}},  \label{delta0}
\end{equation}
where $A$ is the area of the plates.

Thus, from Eqs.~(\ref{vee1}) and (\ref{delta0}) we obtain 
\begin{equation}
E_{0} = -\frac{4A}{(2\pi)^{2}} \sum_{n=1}^{\infty} \int
d^{2}p_{\parallel}\,\omega .  \label{vee1F}
\end{equation}

The overall minus sign is a fermionic feature of the Elko field. The result
is eight times bigger than the scalar-field contribution, or twice that of
the Dirac field. This contrasts with Refs.~\cite%
{maluf2020casimir,Pereira:2016muy}, where the authors report a factor equal
to the Dirac case.

Finally, we emphasize that we neglected the matrix $\mathcal{G}(\phi)$
because our model assumes idealized plates without physical features such as
rugosity. Such features break azimuthal symmetry and induce a preferred
direction in the plane of the plates (see Ref.~\cite{Ahluwalia:2004ab}). For
ideal plates, the symmetry implies that $\mathcal{G}(\phi)$ yields no
contribution.


\bigskip

\begin{thebibliography}{10}

\bibitem{casimir1948attraction}
H.~B. Casimir, \emph{On the attraction between two perfectly conducting
  plates},  in \emph{Proc. Kon. Ned. Akad. Wet.}, vol.~51, p.~793, 1948.

\bibitem{Sparnaay:1958wg}
M.~J. Sparnaay, \emph{{Measurements of attractive forces between flat plates}},
  \href{http://dx.doi.org/10.1016/S0031-8914(58)80090-7}{\emph{Physica}
  {\bfseries 24} (1958) 751--764}.

\bibitem{bressi2002measurement}
G.~Bressi, G.~Carugno, R.~Onofrio and G.~Ruoso, \emph{{Measurement of the
  Casimir force between parallel metallic surfaces}},
  \href{http://dx.doi.org/10.1103/PhysRevLett.88.041804}{\emph{Phys. Rev.
  Lett.} {\bfseries 88} (2002) 041804},
  [\href{https://arxiv.org/abs/quant-ph/0203002}{{\ttfamily
  quant-ph/0203002}}].

\bibitem{lamoreaux1997demonstration}
S.~K. Lamoreaux, \emph{Demonstration of the casimir force in the 0.6 to
  $6\ensuremath{\mu}m$ range},
  \href{http://dx.doi.org/10.1103/PhysRevLett.78.5}{\emph{Phys. Rev. Lett.}
  {\bfseries 78} (Jan, 1997) 5--8}.

\bibitem{lamoreaux1998erratum}
S.~Lamoreaux, \emph{Erratum: Demonstration of the casimir force in the 0.6 to 6
  $\mu$ m range [phys. rev. lett. 78, 5 (1997)]}, {\emph{Physical Review
  Letters} {\bfseries 81} (1998) 5475}.

\bibitem{mohideen1998precision}
U.~Mohideen and A.~Roy, \emph{{Precision measurement of the Casimir force from
  0.1 to 0.9 micrometers}},
  \href{http://dx.doi.org/10.1103/PhysRevLett.81.4549}{\emph{Phys. Rev. Lett.}
  {\bfseries 81} (1998) 4549--4552},
  [\href{https://arxiv.org/abs/physics/9805038}{{\ttfamily physics/9805038}}].

\bibitem{mostepanenko2000new}
V.~M. Mostepanenko, \emph{New experimental results on the casimir effect},
  {\emph{Brazilian Journal of Physics} {\bfseries 30} (2000) 309--315}.

\bibitem{kim2008anomalies}
W.~J. Kim, M.~Brown-Hayes, D.~A.~R. Dalvit, J.~H. Brownell and R.~Onofrio,
  \emph{Anomalies in electrostatic calibrations for the measurement of the
  casimir force in a sphere-plane geometry},
  \href{http://dx.doi.org/10.1103/PhysRevA.78.020101}{\emph{Phys. Rev. A}
  {\bfseries 78} (Aug, 2008) 020101}.

\bibitem{wei2010results}
Q.~Wei, D.~A.~R. Dalvit, F.~C. Lombardo, F.~D. Mazzitelli and R.~Onofrio,
  \emph{Results from electrostatic calibrations for measuring the casimir force
  in the cylinder-plane geometry},
  \href{http://dx.doi.org/10.1103/PhysRevA.81.052115}{\emph{Phys. Rev. A}
  {\bfseries 81} (May, 2010) 052115}.

\bibitem{bordag2001new}
M.~Bordag, U.~Mohideen and V.~M. Mostepanenko, \emph{New developments in the
  casimir effect}, {\emph{Physics reports} {\bfseries 353} (2001) 1--205}.

\bibitem{mostepanenko1997casimir}
V.~Mostepanenko, N.~Trunov and R.~Znajek, \emph{The Casimir Effect and Its
  Applications}.
\newblock Oxford science publications. Clarendon Press, 1997.

\bibitem{brevik2002entropy}
I.~Brevik, K.~A. Milton and S.~D. Odintsov, \emph{Entropy bounds in r{\~u}s3
  geometries}, {\emph{Annals of Physics} {\bfseries 302} (2002) 120--141}.

\bibitem{zhang2015thermal}
A.~Zhang, \emph{Thermal casimir effect in kerr space--time}, {\emph{Nuclear
  Physics B} {\bfseries 898} (2015) 220--228}.

\bibitem{henke2018quantum}
C.~Henke, \emph{Quantum vacuum energy in general relativity}, {\emph{The
  European Physical Journal C} {\bfseries 78} (2018) 1--4}.

\bibitem{milton2019casimir}
K.~Milton and I.~Brevik, \emph{Casimir physics and applications},  2019.

\bibitem{bordag2009advances}
M.~Bordag, G.~L. Klimchitskaya, U.~Mohideen and V.~M. Mostepanenko,
  \emph{Advances in the Casimir Effect}, vol.~145 of \emph{International Series
  of Monographs on Physics}.
\newblock Oxford University Press, Oxford, 2009.

\bibitem{milton2001casimir}
K.~A. Milton, \emph{The Casimir effect: physical manifestations of zero-point
  energy}.
\newblock World Scientific, 2001.

\bibitem{romeo2002casimir}
A.~Romeo and A.~A. Saharian, \emph{{Casimir effect for scalar fields under
  Robin boundary conditions on plates}},
  \href{http://dx.doi.org/10.1088/0305-4470/35/5/312}{\emph{J. Phys. A}
  {\bfseries 35} (2002) 1297--1320},
  [\href{https://arxiv.org/abs/hep-th/0007242}{{\ttfamily hep-th/0007242}}].

\bibitem{aleixo2021thermal}
G.~Aleixo and H.~F.~S. Mota, \emph{{Thermal Casimir effect for the scalar field
  in flat spacetime under a helix boundary condition}},
  \href{http://dx.doi.org/10.1103/PhysRevD.104.045012}{\emph{Phys. Rev. D}
  {\bfseries 104} (2021) 045012},
  [\href{https://arxiv.org/abs/2105.08220}{{\ttfamily 2105.08220}}].

\bibitem{Escobar:2023hzz}
C.~A. Escobar, A.~Mart\'\i{}n-Ruiz, R.~Linares and J.~M. Silva, \emph{{A
  coherent state approach to the Casimir effect for a massive scalar field in a
  noncommutative spacetime}},
  \href{http://dx.doi.org/10.1016/j.aop.2023.169570}{\emph{Annals Phys.}
  {\bfseries 460} (2024) 169570}.

\bibitem{Saghian:2012zy}
R.~Saghian, M.~A. Valuyan, A.~Seyedzahedi and S.~S. Gousheh, \emph{{Casimir
  Energy For a Massive Dirac Field in One Spatial Dimension: A Direct
  Approach}}, \href{http://dx.doi.org/10.1142/S0217751X12500388}{\emph{Int. J.
  Mod. Phys. A} {\bfseries 27} (2012) 1250038},
  [\href{https://arxiv.org/abs/1204.3181}{{\ttfamily 1204.3181}}].

\bibitem{flachi2017sign}
A.~Flachi, M.~Nitta, S.~Takada and R.~Yoshii, \emph{Sign flip in the casimir
  force for interacting fermion systems}, {\emph{Physical review letters}
  {\bfseries 119} (2017) 031601}.

\bibitem{Saharian:2003sp}
A.~A. Saharian and E.~R. Bezerra~de Mello, \emph{{Spinor Casimir densities for
  a spherical shell in the global monopole space-time}},
  \href{http://dx.doi.org/10.1088/0305-4470/37/10/017}{\emph{J. Phys. A}
  {\bfseries 37} (2004) 3543},
  [\href{https://arxiv.org/abs/hep-th/0307261}{{\ttfamily hep-th/0307261}}].

\bibitem{Pereira:2016muy}
S.~H. Pereira, J.~M. Hoff~da Silva and R.~dos Santos, \emph{{Casimir effect for
  Elko fields}}, \href{http://dx.doi.org/10.1142/S0217732317300166}{\emph{Mod.
  Phys. Lett. A} {\bfseries 32} (2017) 1730016},
  [\href{https://arxiv.org/abs/1611.01013}{{\ttfamily 1611.01013}}].

\bibitem{maluf2020casimir}
R.~V. Maluf, D.~M. Dantas and C.~A.~S. Almeida, \emph{{The Casimir effect for
  the scalar and Elko fields in a Lifshitz-like field theory}},
  \href{http://dx.doi.org/10.1140/epjc/s10052-020-8020-9}{\emph{Eur. Phys. J.
  C} {\bfseries 80} (2020) 442},
  [\href{https://arxiv.org/abs/1905.04824}{{\ttfamily 1905.04824}}].

\bibitem{Ahluwalia:2004sz}
D.~V. Ahluwalia and D.~Grumiller, \emph{{Dark matter: A Spin one half fermion
  field with mass dimension one?}},
  \href{http://dx.doi.org/10.1103/PhysRevD.72.067701}{\emph{Phys. Rev. D}
  {\bfseries 72} (2005) 067701},
  [\href{https://arxiv.org/abs/hep-th/0410192}{{\ttfamily hep-th/0410192}}].

\bibitem{Ahluwalia:2004ab}
D.~V. Ahluwalia and D.~Grumiller, \emph{{Spin half fermions with mass dimension
  one: Theory, phenomenology, and dark matter}},
  \href{http://dx.doi.org/10.1088/1475-7516/2005/07/012}{\emph{JCAP} {\bfseries
  07} (2005) 012}, [\href{https://arxiv.org/abs/hep-th/0412080}{{\ttfamily
  hep-th/0412080}}].

\bibitem{jackiw1974functional}
R.~Jackiw, \emph{{Functional evaluation of the effective potential}},
  \href{http://dx.doi.org/10.1103/PhysRevD.9.1686}{\emph{Phys. Rev. D}
  {\bfseries 9} (1974) 1686}.

\bibitem{ryder1996quantum}
L.~H. Ryder, \emph{Quantum field theory}.
\newblock Cambridge university press, 1996.

\bibitem{toms1980symmetry}
D.~J. Toms, \emph{{Symmetry Breaking and Mass Generation by Space-time
  Topology}}, \href{http://dx.doi.org/10.1103/PhysRevD.21.2805}{\emph{Phys.
  Rev. D} {\bfseries 21} (1980) 2805}.

\bibitem{cruz2020casimir}
M.~B. Cruz, E.~R. Bezerra~de Mello and H.~F. Santana~Mota, \emph{{Casimir
  energy and topological mass for a massive scalar field with Lorentz
  violation}}, \href{http://dx.doi.org/10.1103/PhysRevD.102.045006}{\emph{Phys.
  Rev. D} {\bfseries 102} (2020) 045006},
  [\href{https://arxiv.org/abs/2005.09513}{{\ttfamily 2005.09513}}].

\bibitem{porfirio2021ground}
P.~J. Porf\'\i{}rio, H.~F. Santana~Mota and G.~Q. Garcia, \emph{{Ground state
  energy and topological mass in spacetimes with nontrivial topology}},
  \href{http://dx.doi.org/10.1142/S0218271821500565}{\emph{Int. J. Mod. Phys.
  D} {\bfseries 30} (2021) 2150056},
  [\href{https://arxiv.org/abs/1908.00511}{{\ttfamily 1908.00511}}].

\bibitem{PhysRevD.107.125019}
A.~J. D.~F. Junior and H.~F.~S. Mota, \emph{Casimir effect, loop corrections,
  and topological mass generation for interacting real and complex scalar
  fields in minkowski spacetime with different conditions},
  \href{http://dx.doi.org/10.1103/PhysRevD.107.125019}{\emph{Phys. Rev. D}
  {\bfseries 107} (Jun, 2023) 125019}.

\bibitem{toms1980interacting}
D.~J. Toms, \emph{{Interacting Twisted and Untwisted Scalar Fields in a
  Nonsimply Connected Space-time}},
  \href{http://dx.doi.org/10.1016/0003-4916(80)90392-9}{\emph{Annals Phys.}
  {\bfseries 129} (1980) 334}.

\bibitem{PhysRevD.110.045006}
J.~Ven\^ancio, L.~Filho, H.~Mota and A.~Mohammadi, \emph{Thermal casimir effect
  for a dirac field on flat space with a nontrivial circular boundary
  condition}, \href{http://dx.doi.org/10.1103/PhysRevD.110.045006}{\emph{Phys.
  Rev. D} {\bfseries 110} (Aug, 2024) 045006}.

\bibitem{hawking1977zeta}
S.~W. Hawking, \emph{{Zeta Function Regularization of Path Integrals in Curved
  Space-Time}}, \href{http://dx.doi.org/10.1007/BF01626516}{\emph{Commun. Math.
  Phys.} {\bfseries 55} (1977) 133}.

\bibitem{venancio2024thermal}
J.~Venâncio, L.~Filho, H.~Mota and A.~Mohammadi, \emph{Thermal casimir effect
  for a dirac field on flat space with a nontrivial circular boundary
  condition},  2024.

\bibitem{Pereira:2018xyl}
S.~H. Pereira and R.~S. Costa, \emph{{Partition function for a mass dimension
  one fermionic field and the dark matter halo of galaxies}},
  \href{http://dx.doi.org/10.1142/S0217732319501268}{\emph{Mod. Phys. Lett. A}
  {\bfseries 34} (2019) 1950126},
  [\href{https://arxiv.org/abs/1807.06944}{{\ttfamily 1807.06944}}].

\bibitem{doi:10.1142/S021827182450069X}
A.~J.~D. Farias~Junior, A.~Smirnov, H.~F. Santana~Mota and E.~R. Bezerra~de
  Mello, \emph{Vacuum energy density for interacting real and complex scalar
  fields in a lorentz symmetry violation scenario},
  \href{http://dx.doi.org/10.1142/S021827182450069X}{\emph{International
  Journal of Modern Physics D} {\bfseries 34} (2025) 2450069},
  [\href{https://arxiv.org/abs/https://doi.org/10.1142/S021827182450069X}{{\ttfamily
  https://doi.org/10.1142/S021827182450069X}}].

\bibitem{abramowitz1965handbook}
M.~Abramowitz and I.~A. Stegun, \emph{Handbook of mathematical functions dover
  publications}, {\emph{New York} {\bfseries 361} (1965) }.

\bibitem{Elizalde:1995hck}
E.~Elizalde, \emph{Ten Physical Applications of Spectral Zeta Functions},
  vol.~35 of \emph{Lecture Notes in Physics Monographs}.
\newblock Springer-Verlag, Berlin, Heidelberg, 1995,
  \href{http://dx.doi.org/10.1007/978-3-540-44757-3}{10.1007/978-3-540-44757-3}.

\bibitem{Elizalde_hiroshima}
E.~Elizalde, \emph{Explicit analytical continuation of the inhomogeneous
  epstein zeta function}, {\emph{Hiroshima University Preprint} (Sept., 1994)
  }.

\bibitem{Junior:2025thl}
A.~J. D.~F. Junior, A.~Erdas and H.~F. Santana~Mota, \emph{{Vacuum Energy and
  Topological Mass from a Constant Magnetic Field and Boundary Conditions in
  Coupled Scalar Field Theories}},
  \href{https://arxiv.org/abs/2508.15121}{{\ttfamily 2508.15121}}.

\bibitem{elizalde1995zeta}
E.~Elizalde, S.~D. Odintsov, A.~Romeo, A.~A. Bytsenko and S.~Zerbini,
  \emph{{Zeta regularization techniques with applications}}.
\newblock World Scientific Publishing, Singapore, 1994,
  \href{http://dx.doi.org/10.1142/2065}{10.1142/2065}.

\bibitem{Aj}
A.~J.~D. Farias~Junior and H.~F. Mota~Santana, \emph{{Loop correction to the
  scalar Casimir energy density and generation of topological mass due to a
  helix boundary condition in a scenario with Lorentz violation}},
  \href{http://dx.doi.org/10.1142/S0218271822501267}{\emph{Int. J. Mod. Phys.
  D} {\bfseries 31} (2022) 2250126},
  [\href{https://arxiv.org/abs/2204.09400}{{\ttfamily 2204.09400}}].

\bibitem{gradshteyn2014table}
I.~S. Gradshteyn and I.~M. Ryzhik, \emph{Table of integrals, series, and
  products}.
\newblock Academic press, 2014.

\bibitem{ahluwalia2017theory}
D.~V. Ahluwalia, \emph{The theory of local mass dimension one fermions of spin
  one half}, {\emph{Advances in Applied Clifford Algebras} {\bfseries 27}
  (2017) 2247--2285}.

\end{thebibliography}

\end{document}